\documentclass[pdflatex,sn-mathphys-ay]{sn-jnl}% Math and Physical Sciences Author Year Reference Style
\usepackage{graphicx}%
\usepackage{multirow}%
\usepackage{amsmath,amssymb,amsfonts}%
\usepackage{amsthm}%
\usepackage{mathrsfs}%
\usepackage[title]{appendix}%
\usepackage{xcolor}%
\usepackage{textcomp}%
\usepackage{manyfoot}%
\usepackage{booktabs}%
\usepackage{algorithm}%
\usepackage{algorithmicx}%
\usepackage{algpseudocode}%
\usepackage{listings}%
\usepackage{lscape}

\usepackage{subcaption}  % or subfig if you use \subfigure
\usepackage{textcomp}    % for \textsuperscript

\theoremstyle{thmstyleone}%
\theoremstyle{thmstyletwo}%
\newtheorem{remark}{Remark}%

\theoremstyle{thmstylethree}%

\begin{document}

\title[PCA-Guided Quantile Based Sampling]{ Efficient Data Reduction Via PCA-Guided Quantile Based Sampling}


\author[1]{\fnm{Hui-Mean} \sur{Foo}}\email{huimean87@gmail.com}

\author*[1]{\fnm{Yuan-chin Ivan} \sur{Chang}}\email{ycchang@as.edu.tw}
\equalcont{These authors contributed equally to this work.}


\affil*[1]{\orgdiv{Institute of Statistical Science}, \orgname{Academia Sinica}, \orgaddress{\street{128, Academia Road Section 2}, \city{Taipei}, \postcode{11529}, \state{Taiwan}, \country{ROC}}}

%\affil[2]{\orgdiv{Department}, \orgname{Organization}, \orgaddress{\street{Street}, \city{City}, \postcode{10587}, \state{State}, \country{Country}}}

%\affil[3]{\orgdiv{Department}, \orgname{Organization}, \orgaddress{\street{Street}, \city{City}, \postcode{610101}, \state{State}, \country{Country}}}

\abstract{
In large-scale statistical modeling, reducing data size through subsampling is essential for balancing computational efficiency and statistical accuracy. We propose a new method, Principal Component Analysis guided Quantile Sampling (PCA-QS), which projects data onto principal components and applies quantile-based sampling to retain representative and diverse subsets. Compared with uniform random sampling, leverage score sampling, and coreset methods, PCA-QS consistently achieves lower mean squared error and better preservation of key data characteristics, while also being computationally efficient. This approach is adaptable to a variety of data scenarios and shows strong potential for broad applications in statistical computing.
}



\maketitle

\section{Introduction}
\label{sec:intro}

The challenge of analyzing high-dimensional datasets is a fundamental problem in modern data science. As datasets grow in size and complexity, computational burdens increase, often straining conventional analytical techniques that seek to balance efficiency and representativity.

Principal Component Analysis (PCA) is widely used for dimensionality reduction, effectively capturing variance in lower-dimensional representations. Alternative methods, such as variational autoencoders, provide nonlinear transformations to learn latent representations in high-dimensional data \citep{vahdat2020nvae}. However, traditional PCA-based approaches do not explicitly ensure that the reduced dataset remains representative of the original distribution, leading to potential loss of important structural information. Similarly, quantile-based methods provide robust statistical techniques for handling skewed or heterogeneous data but are primarily designed for distribution modeling rather than data reduction. Existing approaches generally focus either on dimensionality reduction without instance selection or on instance selection without leveraging data structure insights. The proposed PCA-QS method bridges this gap by integrating PCA with quantile-based sampling to achieve data reduction in a way that neither technique alone can accomplish.

PCA captures global variance patterns while discarding the original feature space, often at the cost of interpretability \citep{hotelling1933analysis, jolliffe2002principal}. Nonlinear embedding methods such as t-SNE and UMAP are primarily designed for visualization and feature extraction rather than systematic data reduction \citep{becht2019umap}. While they effectively preserve local structure in high-dimensional data, they do not provide a structured framework for selecting representative subsets of instances. Sampling-based approaches such as random and stratified sampling aim to construct representative subsets but may introduce sampling bias or fail to account for the underlying data structure \citep{data_reduction_explained}. More sophisticated methods have been proposed \citep{sarwar2000application, erokhin2022, 9355315, AlKarawi_AlJanabi_2022} to balance efficiency and representativity, yet challenges persist in ensuring that both data integrity and computational feasibility are preserved. These information from literature pointed out the specific challenges but exhibit notable limitations.

Unlike conventional PCA, which transforms variables into principal components and discards the original feature space, PCA-QS leverages PCA to guide instance selection rather than merely reducing dimensionality. Randomized PCA methods have been proposed to improve computational efficiency \citep{halko2011finding}, yet they focus primarily on linear projections rather than instance selection. Quantile-based sampling is then applied in the PCA-transformed space to ensure that the retained subset remains structurally representative of the full dataset. By integrating these two techniques, PCA-QS retains interpretability while significantly improving computational efficiency, overcoming a key limitation of conventional PCA-based dimensionality reduction.

The advantages of PCA-QS extend beyond traditional sampling and dimensionality reduction techniques. By employing PCA as an organizational tool rather than solely as a feature extraction mechanism, PCA-QS ensures that reduced datasets preserve both variance structure and distributional integrity. The quantile-based selection process enables statistically meaningful sampling, capturing the full dataset’s variability and maintaining representativity across key data subgroups. This methodology introduces a novel approach to data reduction that integrates structural insights with representative sampling, offering a practical alternative to conventional methods.
PCA-QS is especially useful when interpretability of original features is required, such as in fields like genomics and social sciences, where understanding variable contributions is critical \citep{hotelling1933analysis, jolliffe2002principal}. In machine learning preprocessing, PCA-QS enhances computational feasibility by enabling model training on compact yet diverse data subsets. Additionally, in big data analytics, it provides an effective strategy for managing large-scale datasets that are otherwise computationally prohibitive to analyze in full.
To evaluate the effectiveness of PCA-QS, we conduct empirical studies using linear regression models as well as clustering applications. These studies assess key performance metrics, including computational efficiency, predictive accuracy, and information retention, by comparing models trained on PCA-QS reduced datasets against other methods. The empirical study also explores PCA-QS in the context of various adaptive sampling strategies, including A-optimal, D-optimal, G-optimal, and uncertainty-based sampling, demonstrating its flexibility across different instance selection frameworks. Furthermore, PCA-QS is applied to widely used large-scale datasets from data archives such as the UCI Machine Learning Repository, illustrating its scalability in real-world scenarios. The results demonstrate that PCA-QS provides a practical and computationally efficient solution for high-dimensional data reduction in both supervised learning and clustering tasks.

The remainder of this paper is structured as follows. Section \ref{sec:method} presents the PCA-QS framework and its implementation. Section \ref{sec:num} discusses empirical evaluations, emphasizing computational efficiency and predictive accuracy. Section \ref{sec:conc} concludes with key findings, practical applications, and future research directions.

\section{Methods} 
\label{sec:method}

{
\subsection{Overview of PCA-Quantile Sampling}

PCA-QS method integrates PCA with quantile-based sampling to achieve efficient data reduction while ensuring representative instance selection. The primary goal is to reduce dataset size while maintaining statistical integrity, enabling efficient modeling in high-dimensional data without losing key structural properties.
The PCA-QS framework consists of the following key steps:
\begin{enumerate}
    \item \textit{PCA Transformation:} Project the data onto the top \( k \) principal components to organize the data structure.
    \item \textit{Quantile-Based Sampling:} Divide instances into quantile groups in PCA-transformed space to ensure balanced representativity.
    \item \textit{Sample Selection:} Extract a subset of instances based on a given deduction rate, preserving the original feature space.
\end{enumerate}

\begin{remark}{Extended PCA for Binary Data} 
While PCA is commonly applied to continuous data, some real datasets used in this study include binary variables. To accommodate this, we employ the {\tt irlba} package in {\it R}, which efficiently computes truncated Singular Value Decomposition (SVD) for large sparse matrices. This package is particularly suited for high-dimensional binary matrices as it does not require explicit dense matrix storage, ensuring computational and memory efficiency \citep{irlba2021, Baglama2005, baglama2019, baglama2024}. Additional randomized decomposition methods are discussed in \citep{erichson2016}, with further applications found in \citep{ai2021, taguchi2021}.
\end{remark}

\subsection*{Quantile-Based Sampling in PCA Space}
Instead of randomly selecting instances, PCA-QS uses quantile-based stratification to preserve dataset representativity. The key idea is:
(a) partition data into quantile-based groups along each principal component, and (b) ensure that all parts of the distribution are evenly represented.

\paragraph{Principal Component Analysis as a Structural Guide}
PCA is a fundamental dimensionality reduction technique that projects high-dimensional data onto a lower-dimensional space that captures maximal variance \citep{hotelling1933analysis, jolliffe2002principal}. Given a dataset $\mathbf{X} \in \mathbb{R}^{n \times p}$ with $n$ samples and $p$ features, PCA is performed using Singular Value Decomposition (SVD):
\begin{equation}
    \mathbf{X} = \mathbf{U} \mathbf{D} \mathbf{V}^\top,
\end{equation}
where:
$\mathbf{U} \in \mathbb{R}^{n \times p}$ contains the left singular vectors (principal components),
$\mathbf{D} \in \mathbb{R}^{p \times p}$ is a diagonal matrix of singular values,
$\mathbf{V} \in \mathbb{R}^{p \times p}$ contains the right singular vectors (principal axes).  
The data is transformed into the PCA space as:
\begin{equation}
    \mathbf{Z} = \mathbf{U}_k \mathbf{D}_k,
\end{equation}
where $\mathbf{Z} \in \mathbb{R}^{n \times k}$ represents the dataset in the new reduced space using the top \( k \) principal components that capture the majority of variance.
Although PCA captures data variance, it does not perform instance selection, meaning that datasets remain large and computationally expensive to analyze, and that is why PCA alone is not enough. PCA-QS bridges this gap by leveraging PCA for structured sampling.

\paragraph{PCA Transformed Space for \( \text{num\_pcs} = 1 \)}
To divide data into $g$ quantile groups $(g \in N)$, we define the quantile stratification as follows:
\begin{equation}\label{eq:pca1}
    \left[\min, Q_1\right), \left[Q_1, Q_2\right), \left[Q_2, Q_3\right), \left[Q_3, Q_4\right),  \ldots, \left[Q_{g-1} \max\right],
\end{equation}
where \( Q_i \) represents the \( i \)-th quantile of the PCA-transformed dataset.
The final reduced dataset is constructed by sampling a fixed proportion from each quantile group,  ensuring representativeness while significantly reducing computational complexity.

Once PCA transformation is performed, quantile-based stratification is applied to structure the data. When \( \text{num\_pcs} = 1 \) with $g=5$,  the method partitions the data along the first principal component into quantile-based groups, as in \eqref{eq:pca1}:
\begin{equation}
[\text{min}, Q_1), [Q_1, Q_2), [Q_2, Q_3), [Q_3, Q_4), [Q_4, \text{max}]
\end{equation}
where \( Q_i \) represents the \( i \)-th quantile. This division ensures that each quantile group contains approximately 20\% of the samples, promoting balanced and representative sampling.

\paragraph{PCA Transformed Space for \( \text{num\_pcs} > 1 \)}
When more than one principal component is retained (\( \text{num\_pcs} > 1 \)), the PCA-transformed space becomes a multidimensional subspace. Quantile-based sampling is extended beyond a single axis, requiring a structured partitioning strategy to ensure that the reduced dataset maintains representativeness across multiple principal components.

\paragraph{Quantile Assignment for Multiple Components}
For \( k > 1 \), each principal component defines an axis in the transformed space, and quantiles are assigned independently for each component. When $g=5$, the quantile assignment follows these steps:
\begin{enumerate}
    \item \textbf{Component-wise Quantiles:}  
    Each principal component \( z_j \) is partitioned into quantiles based on its distribution, analogous to the 
	\( \text{num\_pcs} = 1 \) case. For example, with five quantile partitions:
    \[
    \text{Quantile Groups for Component } j: [\text{Min}, Q_1), [Q_1, Q_2), \dots, [Q_4, \text{Max}].
    \]
    
    \item \textbf{Composite Quantile Groups:}  
    Each sample is assigned a composite quantile group based on its quantile memberships across all \( k \) components:
    \[
    \text{Quantile Group: } \{q_{i1}, q_{i2}, \ldots, q_{ik}\},
    \]
    where \( q_{ij} \) represents the quantile group of sample \( i \) for component \( j \).
    
    \item \textbf{Unique Group Identifiers:}  
The composite quantile group is encoded as a unique identifier (e.g., "2-4-1" for \( k = 3 \)), where each digit represents the sample’s quantile assignment for the respective principal component.
\end{enumerate}

\subsubsection*{Sampling from Quantile Groups}
To maintain balanced representation, samples are randomly selected from each quantile group. Random selection prevents deterministic biases while ensuring proportional representation across quantile-defined subgroups. The number of selected samples per group is determined by:
\[
\text{Sample Size per Group} = \min\left(\lceil \text{retention rate} \times N_g \rceil, N_g\right),
\]
where \( N_g \) is the number of samples in group \( g \), and the \textit{retention rate} (e.g., \( 0.05 \)) controls the fraction of data retained. This approach preserves structural diversity while reducing computational complexity. 

While random sampling ensures fairness, alternative selection strategies can further optimize sample representativeness. Other selection strategies, such as sampling based on experimental design criteria (e.g., A- and D-optimal designs) and uncertainty-based sampling, can be incorporated within the PCA-QS framework.  In the following numerical study (Section \ref{subsec:Adaptive sampling}), we illustrate how adaptive sampling strategies can be effectively integrated into PCA-QS.

\subsubsection*{Illustrative Example}
Consider a dataset with \( n = 10,000 \) samples and \( p = 50 \) features, where PCA is applied with \( k = 2 \) components. Suppose the retention rate is set to \( \delta = 0.05 \). The PCA-QS procedure follows these steps:

\begin{enumerate}
    \item PCA transforms the original dataset into a lower-dimensional representation, yielding a \( 10,000 \times 2 \) matrix \( \mathbf{Z} \), where each row represents a sample in the 2 dimensional principal component space.
    
    \item Each principal component is divided into 5 quantiles. If sample \( i \) belongs to the 3rd quantile of the first principal component and the 5th quantile of the second, it is assigned to composite quantile group \( (3,5) \).
    
    \item Since there are \( 5^2 = 25 \) total composite quantile groups, each sample is assigned to exactly one group.
    
    \item Sampling: If quantile group \( (3,5) \) contains \( N_{3,5} = 400 \) samples, the retained sample size for this group is:
    \(
    \textbf{Samples retained} = \lceil 0.05 \times 400 \rceil = 20.
    \)
    
    \item This procedure is applied across all \( 25 \) composite groups to form the final retentive dataset, ensuring representativeness across principal component space.
\end{enumerate}

\subsubsection*{Algorithmic Implementation of PCA-QS}
To provide a clear computational framework, we present the PCA-QS algorithm.
\begin{algorithm}[h!]
\caption{PCA-Quantile Sampling (PCA-QS) Framework}
\begin{algorithmic}[1]
    \State \textbf{Input:} Dataset \( \mathbf{X} \in \mathbb{R}^{n \times p} \) with \( n \) samples and \( p \) features.
    \State \textbf{Preprocessing:} Standardize features to zero mean and unit variance.
    \State \textbf{Step 1:} Apply PCA to transform \( \mathbf{X} \) into the principal component space.
    \State \textbf{Step 2:} Retain the top \( k \) principal components that capture sufficient variance.
    \State \textbf{Step 3:} Assign each sample to quantile groups based on its transformed PCA coordinates.
    \State \textbf{Step 4:} Select representative instances from each quantile group based on a given \textbf{retention rate} \( \delta \).
    \State \textbf{Step 5:} Construct the final \textbf{retentive dataset} \( \mathbf{X}_{\text{retentive}} \) for downstream analysis.
    \State \textbf{Output:} Retentive dataset \( \mathbf{X}_{\text{retentive}} \) preserving representative structure.
\end{algorithmic}
\end{algorithm}

\subsection{Mathematical Properties  of PCA-QS}

Let \( X \in \mathbb{R}^{n \times p} \) be a multivariate normal random matrix:
\(
X \sim \mathcal{N}(0, \Sigma),
\)
where \( \Sigma \) is the population covariance matrix. PCA transforms the dataset into an orthogonal basis by computing the eigenvectors of the sample covariance matrix:
\begin{equation}
S = \frac{1}{n} X^T X.
\end{equation}
The eigenvectors of \( S \) define the principal component (PC) directions, while the corresponding eigenvalues indicate their variance contributions. The PCA transformation is given by:
\(
Z = X V,
\)
where \( V \) is the eigenvector matrix of \( \Sigma \). The first principal component follows:
\(
Z_1 \sim \mathcal{N}(0, \lambda_1),
\)
where \( \lambda_1 \) is the largest eigenvalue.

\paragraph{PCA and Quantile-Based Sampling} 
PCA naturally transforms the data into an uncorrelated basis, allowing for accurate quantile sampling. Even if the original features are correlated (\( \Sigma \neq I \)), PCA ensures that the transformed principal components remain uncorrelated. This justifies performing quantile-based stratification in the PCA space, as each PC can be treated as an independent normal variable:
\begin{equation}
    Z_j \sim \mathcal{N}(0, \lambda_j).
\end{equation}
Thus, PCA-based quantile sampling remains robust regardless of feature dependencies.

\begin{remark}
Applying PCA directly diagonalizes the covariance matrix in the transformed space, making any explicit diagonalization step unnecessary. If the original covariance matrix is already diagonal, PCA only rescales features, and PCA-QS operates directly on these rescaled variables. Artificial diagonalization removes meaningful relationships among features and reduces PCA’s effectiveness.
\end{remark}

\paragraph{Distribution of Top Principal Components}
In a low-dimensional setting (\( p \ll n \)), the top \( k \) principal components follow a multivariate normal distribution:
\(
(Z_1, Z_2, ..., Z_k) \sim \mathcal{N} (0, \Lambda_k),
\)
where \( \Lambda_k = \text{diag}(\lambda_1, \lambda_2, ..., \lambda_k) \) is the diagonal matrix of the top eigenvalues.

Since each principal component follows a normal distribution:
\(
PC_j \sim \mathcal{N}(0, \lambda_j),
\)
where \( \lambda_j \) is the eigenvalue corresponding to PC \( j \), 
and its quantile distribution is:
\begin{equation}
    Q_j(p) = \Phi^{-1}(p) \cdot \sqrt{\lambda_j},
\end{equation}
where \( \Phi^{-1}(p) \) is the standard normal quantile function.

\paragraph{Quantile Sampling with the Top \( k \) Principal Components}
When using the top \( k \) PCs for quantile-based stratification, each principal component's quantiles are scaled by its eigenvalue. If the top 5 PCs are used for sampling, the quantile distributions follow:
\begin{equation}
Q_j(p) = \Phi^{-1}(p) \cdot \sqrt{\lambda_j}, \quad j = 1, ..., 5.
\end{equation}
This ensures that the PCA-QS method selects samples based on variance-scaled normal quantiles, preserving the original distributional structure.

\paragraph{Example: Theoretical Quantile Distribution of Top 5 PCs}

Using 5 quantile groups, we take:
\[
(0\% \text{ (min)}, 20\% (Q_1), 40\% (Q_2), 60\% (Q_3), 80\% (Q_4), 100\% \text{ (max)}).
\]
For a standard normal distribution, these correspond to:
\[
\Phi^{-1}(0.2) = -0.84, \quad
\Phi^{-1}(0.4) = -0.25, \quad
\Phi^{-1}(0.6) = 0.25, \quad
\Phi^{-1}(0.8) = 0.84.
\]
These values are then scaled by \( \sqrt{\lambda_j} \), leading to the following empirical quantiles:

\begin{table}[h]
    \centering
    \begin{tabular}{lrrrrr}
        \toprule
        Quantile & PC1 & PC2 & PC3 & PC4 & PC5 \\
        \midrule
        20\%  & $-0.84 \sqrt{\lambda_1}$ & $-0.84 \sqrt{\lambda_2}$ & $-0.84 \sqrt{\lambda_3}$ & $-0.84 \sqrt{\lambda_4}$ & $-0.84 \sqrt{\lambda_5}$ \\
        40\%  & $-0.25 \sqrt{\lambda_1}$ & $-0.25 \sqrt{\lambda_2}$ & $-0.25 \sqrt{\lambda_3}$ & $-0.25 \sqrt{\lambda_4}$ & $-0.25 \sqrt{\lambda_5}$ \\
        60\%  & $0.25 \sqrt{\lambda_1}$ & $0.25 \sqrt{\lambda_2}$ & $0.25 \sqrt{\lambda_3}$ & $0.25 \sqrt{\lambda_4}$ & $0.25 \sqrt{\lambda_5}$ \\
        80\%  & $0.84 \sqrt{\lambda_1}$ & $0.84 \sqrt{\lambda_2}$ & $0.84 \sqrt{\lambda_3}$ & $0.84 \sqrt{\lambda_4}$ & $0.84 \sqrt{\lambda_5}$ \\
        \bottomrule
    \end{tabular}
    \caption{Quantile Values for Top 5 Principal Components}
\end{table}

If \( \lambda_j \) is small, the quantile values remain close to zero, indicating that the corresponding principal component does not capture meaningful variability in the data. Consequently, quantile-based sampling from such components may introduce bias and fail to preserve the original data structure. To mitigate this, classical methods for determining the number of retained principal components ($k$) can be applied to PCA-QS. One common approach is the cumulative variance threshold, where components are retained until they account for a sufficient proportion of the total variance, typically 95\%. 
Another widely used technique is scree plot analysis, which visually identifies the point where the variance explained by additional components begins to level off. 
Additionally, cross-validation error minimization provides a data-driven criterion for selecting an optimal number of components based on predictive performance. Alternatively, statistical hypothesis testing, such as the broken stick method, can be employed to determine the appropriate dimensionality by assessing whether each principal component explains more variance than expected under a random model.

\begin{remark}[Topic for Future Investigation: High-Dimensional \( p \sim n \) Case]
In very high-dimensional settings where \( p \) grows at the same rate as \( n \), classical PCA assumptions no longer hold. Specifically, the largest eigenvalues of the sample covariance matrix no longer follow a normal distribution but instead exhibit non-Gaussian fluctuations. In such cases, the largest eigenvalue follows the Tracy-Widom distribution, which governs the extreme eigenvalues of large random matrices \citep{forrester2010log}:
\begin{equation}
  \lambda_1 \approx \mu_{\lambda} + n^{-2/3} \cdot TW_1,
\end{equation}
where \( TW_1 \) is the Tracy-Widom distribution (Gaussian Orthogonal Ensemble case) \citep{tracy1994level, tracy1996orthogonal}, and \( \mu_{\lambda} \) is derived from the Marchenko-Pastur theorem \citep{marchenko1967distribution}.
This result is a cornerstone of random matrix theory, which characterizes eigenvalue behavior in large covariance matrices \citep{mehta2004random, baik1999distribution, johnstone2001distribution}. The Tracy-Widom law has significant implications in statistical signal processing, wireless communications, and genomics \citep{baik1999distribution}. 
Although we did not extend PCA-QS to the high-dimensional regime 
\( p \sim n \)  in this study, the method could be adapted for such settings and would make a good topic for future investigation.
\end{remark}
}

\subsubsection*{Computing Advantages and Challenges}
PCA-QS is computationally efficient, as it focuses only on the top few principal components, reducing both memory and computation costs. It is particularly well-suited for large-scale data analysis and can be implemented efficiently within a parallel or distributed computing framework \citep{halko2011finding, martinsson2020randomized, erichson2016, liu2017parallel}.  

In many applications, PCA is commonly applied to continuous data. However, some real-world datasets contain binary variables, requiring modifications to standard PCA. Several R packages offer solutions for handling such cases. For example, the \texttt{irlba} package performs truncated Singular Value Decomposition (SVD) and is particularly effective for binary matrices, as it avoids explicit storage of large matrices, improving both memory efficiency and computational performance \citep{irlba2021}. See also \cite{Baglama2005, baglama2019, baglama2024} for more details on matrix factorization techniques.
Additionally, randomized decomposition methods provide efficient approximations for large-scale problems, as discussed in \citep{erichson2016}. Further applications of these techniques can be found in \citep{ai2021, taguchi2021}.

%\paragraph{Comparison with Existing Methods}
To highlight the advantages of PCA-QS, we compare it with widely used {data reduction} methods based on existing literature including {Random Sampling} \citep{Provost1998, Shao2003}, {Stratified Sampling} \citep{Shao2003}, {Standard PCA} \citep{jolliffe2002principal, Shlens2014}, {Kernel PCA} \citep{scholkopf1998nonlinear, Mika1999, chang2008kernel, Li2009kernel}, {t-SNE/UMAP} \citep{Maaten2008, McInnes2018, Linderman2017, Belkin2001}, {Leverage Score Sampling} \citep{Drineas2006, Mahoney2011} and {Coreset Sampling} \citep{HarPeled2004, feldman2011}.   
Unlike {random sampling}, which may discard critical data structure, or {t-SNE/UMAP}, which is computationally expensive and lacks interpretability, {PCA-QS efficiently preserves both data structure and original features while scaling well to large datasets}.  That is, PCA-QS is a tool for data deduction, which produces a representable retentive subdata set for downstream analysis. Since all features are retained in its subdata set, the interpretability of the final results depends on the follow-up analysis tools used with the retentive data set.  
PCA-QS achieves a strong balance between {computational efficiency, interpretability, and scalability} compared to these traditional methods. 
The following table summarizes the key comparisons:
\begin{table}[h!]
    \centering
    \begin{tabular}{lccc}
        \toprule
        \textbf{Method} & \textbf{Computational Efficiency} & \textbf{Interpretability} & \textbf{Scalability} \\
        \midrule
         PCA-QS (Proposed) & High & High & High \\
        Random Sampling & High & Low & High \\
        Stratified Sampling & Medium & High & Medium \\
        Leverage Score Sampling & High & Medium & High \\
        Coreset Sampling & Low & Medium & Medium \\
        Standard PCA & Medium & Low & High \\
        Kernel PCA & Low & Very Low & Low \\
        t-SNE/UMAP & Low & Very Low & Low \\
        \bottomrule
    \end{tabular}
    \caption{Comparison of Computational Efficiency, Interpretability and Scalability of PCA-QS with some dimension deduction and sampling  methods.}
    \label{tab:method_comparison}
\end{table}

Although, Table \ref{tab:method_comparison} includes t-SNE, UMAP, and Kernel PCA, it is important to note that t-SNE, UMAP, and Kernel PCA are primarily designed for dimensionality reduction, not data reduction. Thus, a direct comparison may not be appropriate, since their objectives are fundamentally different from PCA-QS. However, there is potential to {integrate these nonlinear embedding techniques within the PCA-QS framework in future research} to further enhance instance selection and representation.

\subsection{Practical Considerations: Challenges and Trade-offs}
While PCA-QS provides several advantages, there are important trade-offs to consider. One key challenge is the exponential growth in the number of quantile groups. With 5 quantiles per component and \( k \) retained components, the total number of groups is \( 5^k \). For instance, when \( k = 3 \), there are 125 possible groups, which can significantly increase computational costs. Additionally, some quantile groups may contain very few or no samples, particularly in small datasets or when \( k \) is large. This sparsity can reduce the statistical stability of models trained on the sampled dataset.

Furthermore, while increasing \( k \) retains more variance and improves representativeness, it also raises computational overhead. Selecting an appropriate balance between accuracy and efficiency is crucial for effective data reduction. Another challenge arises when data distributions along PCA components are highly skewed, leading to imbalanced quantile groups. In such cases, additional adjustments may be necessary to ensure that sampling remains representative across different subgroups.

When applying PCA-QS, three key factors should be carefully evaluated: the number of retained PCA components (\( k \)), the optimal retention rate (\( \delta \)), and strategies for handling sparse quantile groups. A smaller \( k \) reduces computational costs but may fail to retain essential variance, potentially degrading model performance. Conversely, a larger \( k \) enhances representativeness but increases the number of quantile groups, potentially leading to sparsity. The conventional approach of selecting \( k \) based on cumulative explained variance remains a useful guideline.

The retention rate should balance computational efficiency with predictive accuracy. Typical values range from 0.01 to 0.1, depending on dataset size and dimensionality. If some quantile groups contain very few samples, fallback strategies such as oversampling or merging adjacent groups can be implemented to maintain representativeness. The choice of these parameters depends on the dataset's characteristics. The numerical studies presented in the following sections provide further insights into these considerations and offer practical guidance for selecting suitable parameter values.

\begin{remark}[Historical and Statistical Foundations]
The statistical properties of PCA and quantile-based methods have been extensively studied in prior research. PCA, originally introduced by \citet{hotelling1933analysis}, has well-documented theoretical properties, including consistency and asymptotic efficiency under normality assumptions \citep{jolliffe2002principal}. It has been widely applied in data science for dimensionality reduction, feature extraction, and noise filtering, with notable applications in genomics \citep{ringner2008pca}, image recognition \citep{turk1991face}, and recommendation systems \citep{sarwar2000application}.

Similarly, quantile-based methods, dating back to \citet{galton1874percentiles}, have been extensively used for modeling skewed or heteroscedastic data. Their robustness properties, such as invariance under monotonic transformations and resilience to outliers, make them valuable tools in statistical inference and predictive modeling \citep{hampel1986robust, koenker1978quantile}. These foundational properties underscore the reliability of PCA-QS as a hybrid approach that leverages the strengths of both methods.
\end{remark}

\section{Numerical Studies}
\label{sec:num}
\subsection{Comparing with Popular Sampling Methods}
\paragraph{Data Generation}

Synthetic data were generated as follows. A mixture of Gaussian distributions was constructed to represent the covariate space. Specifically, the number of samples was set to $n = 10{,}000$ and the number of features to $p = 50$. 
A total of $K$ components were defined, each with a mean vector sampled from a standard normal distribution scaled by a factor of five and an identity covariance matrix:
\[
\boldsymbol{\mu}_k \sim \mathcal{N}(\mathbf{0}, 5^2 \mathbf{I}_p), \quad
\boldsymbol{\Sigma}_k = \mathbf{I}_p.
\]
The mixture weights were drawn from a symmetric Dirichlet distribution:
\[
\boldsymbol{\pi} \sim \text{Dirichlet}(\mathbf{1}_K).
\]
Component assignments for each observation were sampled according to these weights, and the features were generated as:
\[
\mathbf{x}_i \sim \sum_{k=1}^K \pi_k \mathcal{N}(\boldsymbol{\mu}_k, \boldsymbol{\Sigma}_k).
\]
A response variable was then generated by drawing a coefficient vector $\boldsymbol{\beta} \sim \mathcal{N}(\mathbf{0}, \mathbf{I}_p)$, and computing:
\[
y_i = \mathbf{x}_i^\top \boldsymbol{\beta} + \varepsilon_i, 
\quad \varepsilon_i \sim \mathcal{N}(0, 1).
\]
(The random seed was fixed at 42 to ensure reproducibility.)

\begin{figure}[htbp]
  \centering
  \includegraphics[width=0.9\textwidth]{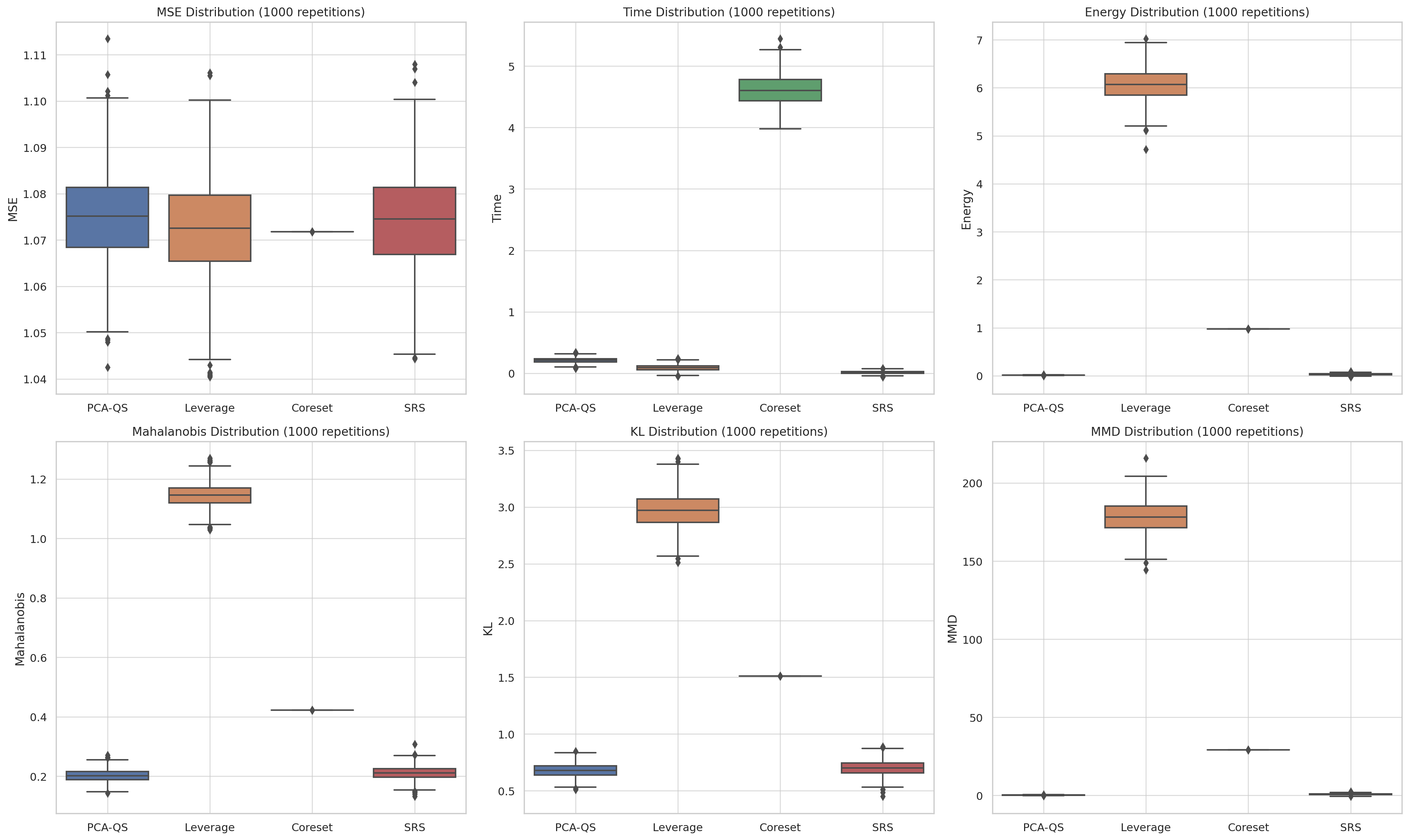}
  \caption{Comparison the ``distances'' between the retentive data sets of sampling methods to the original over data set generated with a gaussian-mixture data. Results shown here is based on 1000 replications under 10\% sampling rate.}
  \label{fig:sampling_comparison}
\end{figure}

Before we focus on the detailed performance of PCA-QS method, we conduct a brief numerical study to compare the retentive data sets obtained with Leverage Score Sampling \citep{Drineas2006, Mahoney2011}, Coreset Sampling \citep{HarPeled2004, feldman2011}, and SRS.
Table~\ref{tab:sampling_comparison} summarizes the performance of these four data reduction techniques, evaluated over 1000 replications on linear regression models based on a synthetic mixture of Gaussians, with 100000 samples, and 50 features, using 10\% samples of the original data.  From this table, we found that 
{PCA-QS} consistently demonstrates a well-balanced trade-off between distributional fidelity and model accuracy. It achieves the lowest energy distance, Mahalanobis distance, KL divergence, and MMD among all methods, indicating superior preservation of the overall data structure. Its mean squared error (MSE) is slightly higher than Leverage and Coreset but remains competitive. The runtime is moderate due to PCA decomposition and quantile stratification.
{Leverage Score Sampling} \citep{Drineas2006, Mahoney2011} yields the lowest MSE next to Coreset, underscoring its strength in retaining influential data points critical for linear model fitting. However, its high energy distance, Mahalanobis distance, KL divergence, and exceptionally large MMD highlight a significant distortion of the original data distribution, as it preferentially samples extreme leverage points.
{Coreset Sampling} \citep{HarPeled2004, feldman2011} achieves the lowest MSE with zero variance, reflecting its stability in summarizing cluster means. Nonetheless, it performs poorly on preserving local variation, as evidenced by higher distributional distances and the largest runtime due to the iterative nature of clustering.
{SRS} serves as a robust baseline, achieving fast execution and fair performance across all metrics. It does not exploit any structural information but provides an unbiased subsample that approximates the original distribution moderately well.
(Figure~\ref{fig:sampling_comparison} presents the box-plots for each metric used in this study, corroborating the numerical findings. The Python code used to generate these results is available upon request for reproducibility.)

\paragraph{Practical Recommendation} Since these methods are developed with different considerations and for different purposes.  Leverage picks rows with extreme leverage, which can be biased distribution but good for linear model estimation.
PCA-QS intentionally spreads out selection in reduced space, so it matches distribution better.
Coreset uses cluster centers, which is great for low error in cluster-wise means, but bad for detailed spread.
SRS just samples uniformly, which is usually unbiased but not optimized for either task. Thus, in practice, PCA-QS is the most suitable when balanced preservation of the global data distribution and interpretability are desired. Leverage sampling excels for tasks focusing purely on predictive accuracy in linear models but at the expense of distributional similarity. Coreset sampling is best for cluster-based summarization, where signal approximation outweighs the need for distributional detail. SRS remains a practical fallback when computational simplicity is paramount.

\begin{table}[htbp]
\centering
\caption{Performance comparison of sampling methods over 1000 replications. Each mean is followed by its standard deviation in parentheses. The smallest (best) value for each metric is shown in bold.}
\begin{tabular}{lrrrrr}
\hline
\textbf{Methods} & \textbf{PCA-QS} & \textbf{Leverage} & \textbf{Coreset} & \textbf{SRS} \\
\hline
MSE & 1.0749 (0.0100) & \textbf{1.0719} (0.0107) & \textbf{1.0718} ($\approx$0) & 1.0745 (0.0103) \\
Execution Time(s) & 0.211 (0.041) & 0.092 (0.049) & 4.615 (0.235) & \textbf{0.015} (0.020) \\
Energy Dist. & \textbf{0.0138} (0.0031) & 6.0695 (0.344) & 0.9771 ($\approx$0) & 0.0325 (0.0164) \\
Mahalanobis Dist. & \textbf{0.2020} (0.0210) & 1.1431 (0.0369) & 0.4229 ($\approx$0) & 0.2111 (0.0216) \\
KL Div. & \textbf{0.6781} (0.0573) & 2.9623 (0.1508) & 1.5117 ($\approx$0) & 0.6991 (0.0643) \\
MMD & \textbf{0.1785} (0.0890) & 177.9425 (10.493) & 29.2147 ($\approx$0) & 0.7427 (0.4920) \\
\hline
\end{tabular}
\label{tab:sampling_comparison}
\end{table}

\subsubsection{Linear Model Example}
To evaluate the performance of PCA-QS for linear models, we generate a synthetic dataset with a large sample size and high feature dimensionality, and compare results obtained from PCA-QS sampling with those based on the full dataset.

Specifically, we generate $n = 10{,}000$ samples with $p = 500$ features using the \texttt{mvrnorm} function from the \texttt{MASS} package. The covariate matrix $X$ follows a multivariate normal distribution with zero mean and an equi-correlated covariance matrix $\Sigma$. The diagonal entries of $\Sigma$ are set to 1 (unit variance), and all off-diagonal entries are set to a constant correlation $\rho = 0.2$:

\[
\Sigma =
\begin{bmatrix}
1 & \rho & \rho & \dots & \rho \\
\rho & 1 & \rho & \dots & \rho \\
\rho & \rho & 1 & \dots & \rho \\
\vdots & \vdots & \vdots & \ddots & \vdots \\
\rho & \rho & \rho & \dots & 1
\end{bmatrix}
\]

The regression coefficient vector $\beta$ is generated from a normal distribution $N(0, 0.1^2)$, and the response variable is defined as:
\[
Y = X \beta + \varepsilon, \quad \text{where } \varepsilon \sim N(0, 0.5^2).
\]

This controlled setting enables an objective assessment of how well PCA-QS retains information relevant for linear regression analysis.

\paragraph{Computational Efficiency and Predictive Performance}

\begin{figure}[htbp]
    \centering
    \subfloat[MSE]{\includegraphics[width=0.45\textwidth]{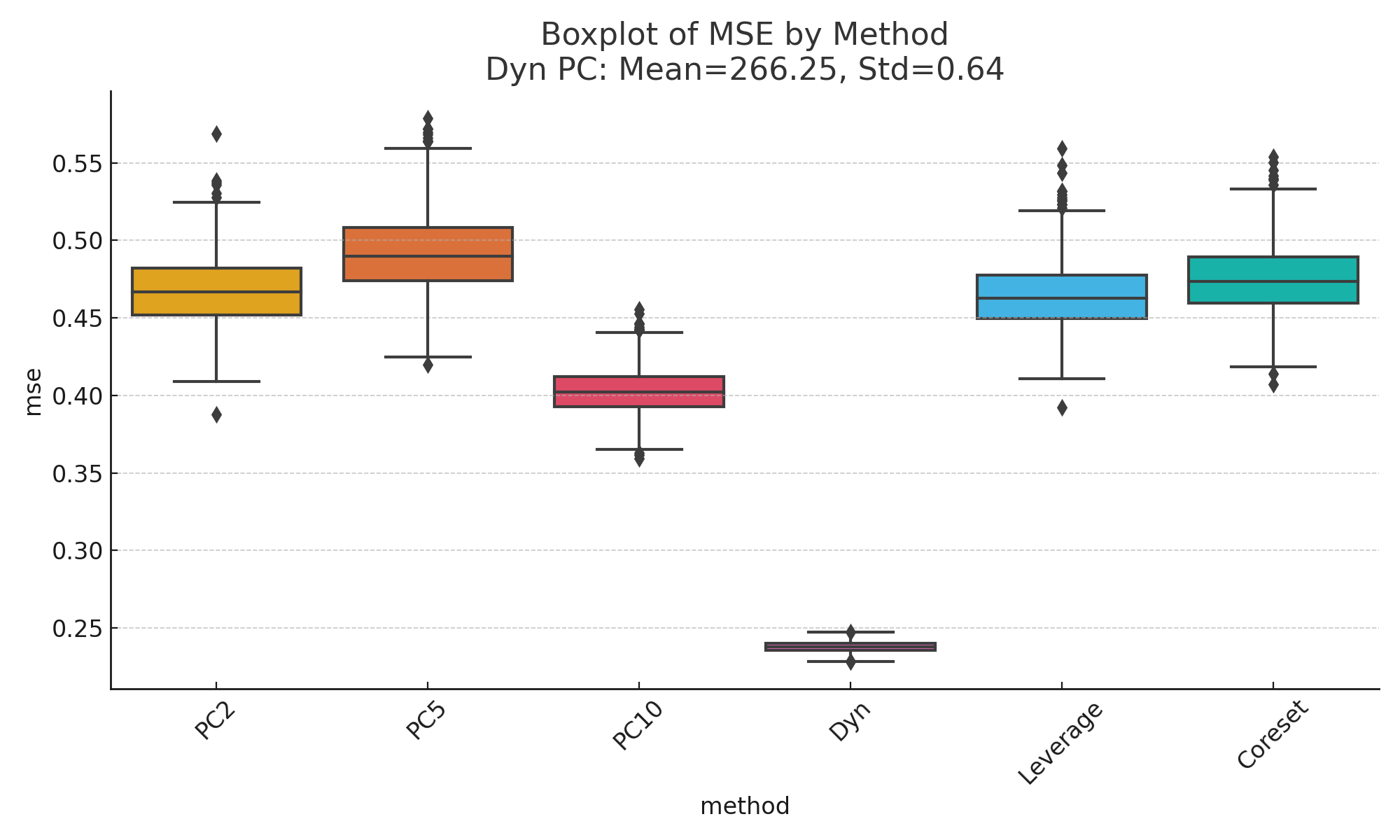}}
    \hfill
    \subfloat[R\textsuperscript{2}]{\includegraphics[width=0.45\textwidth]{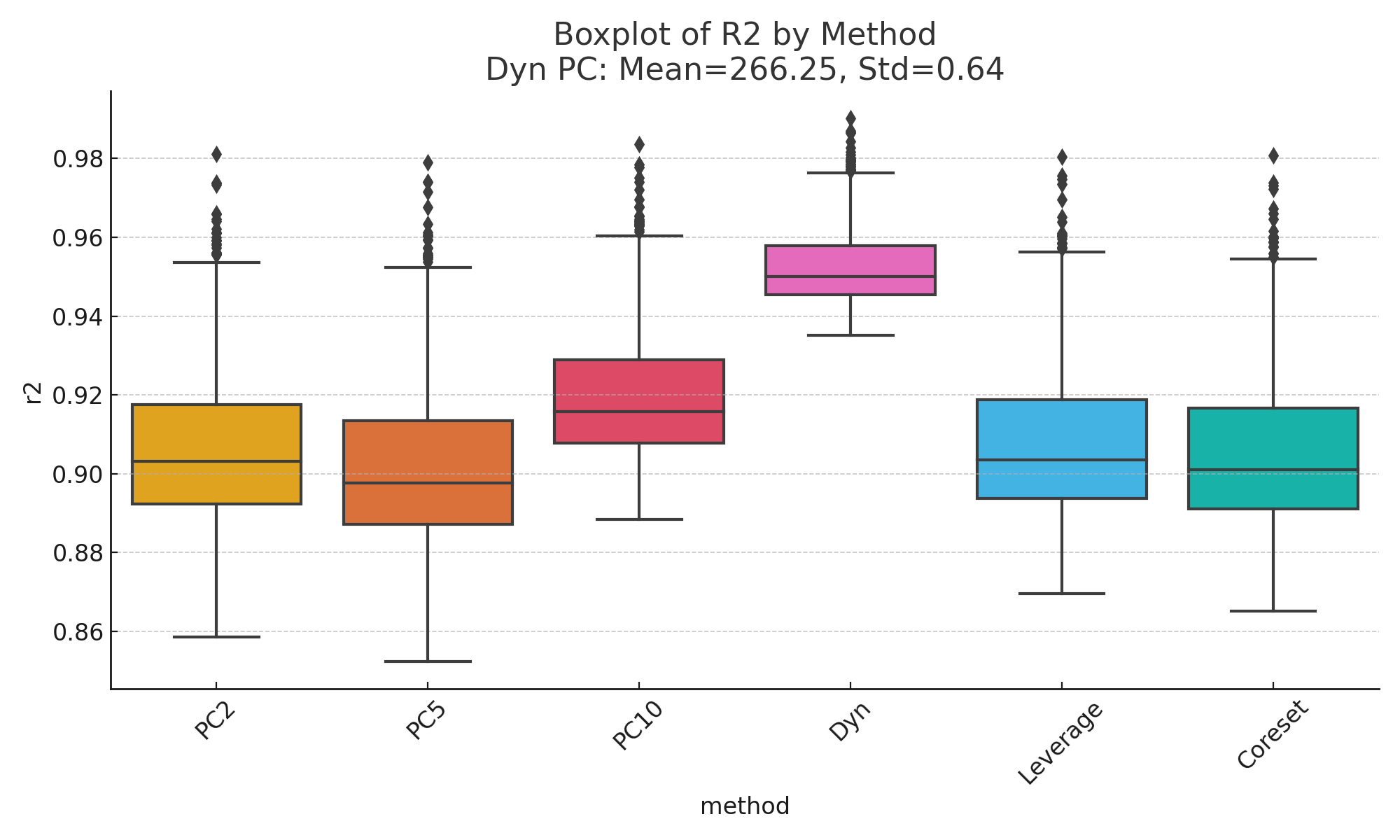}} \\
    \subfloat[Runtime (sec)]{\includegraphics[width=0.45\textwidth]{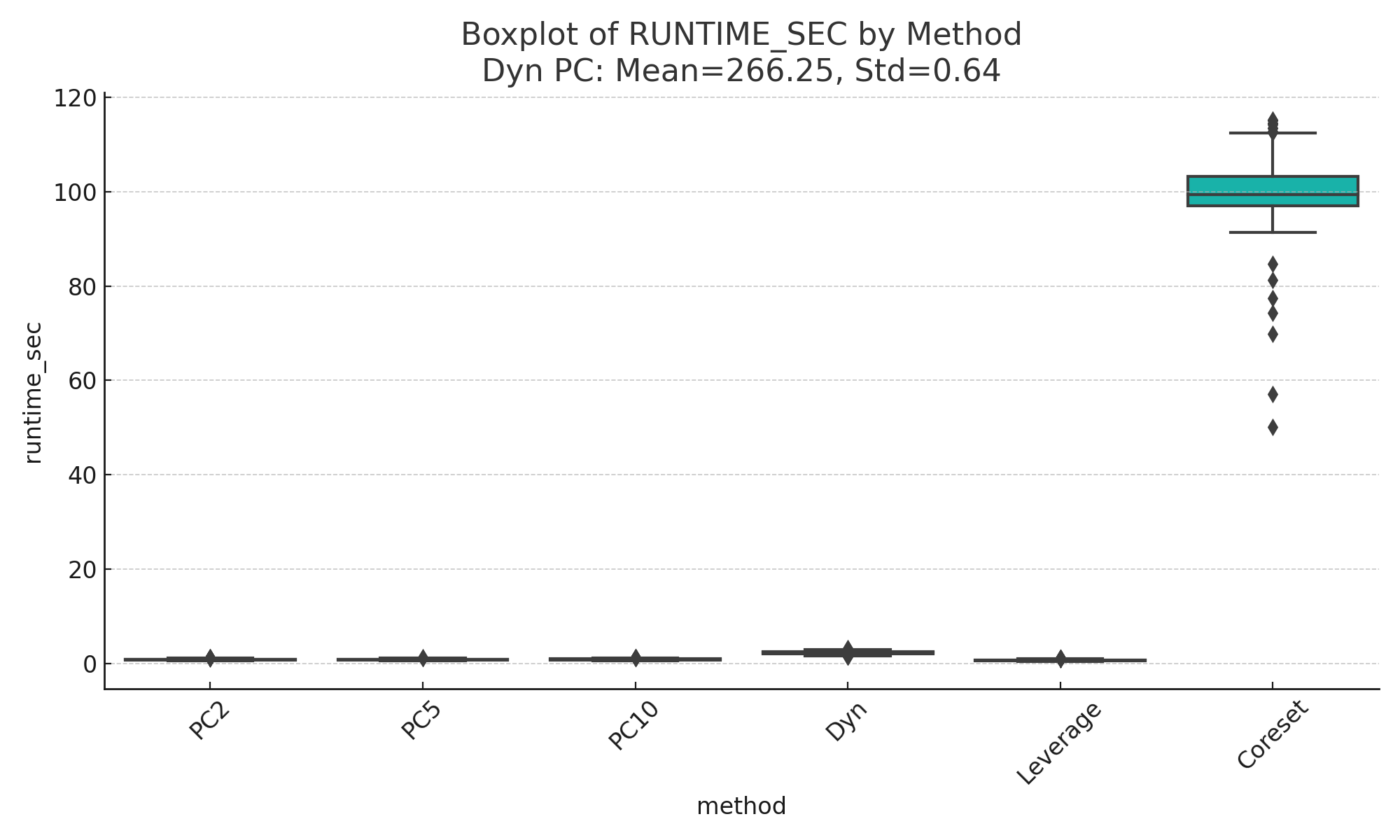}}
    \hfill
    \subfloat[Runtime (excluding Coreset)]{\includegraphics[width=0.45\textwidth]{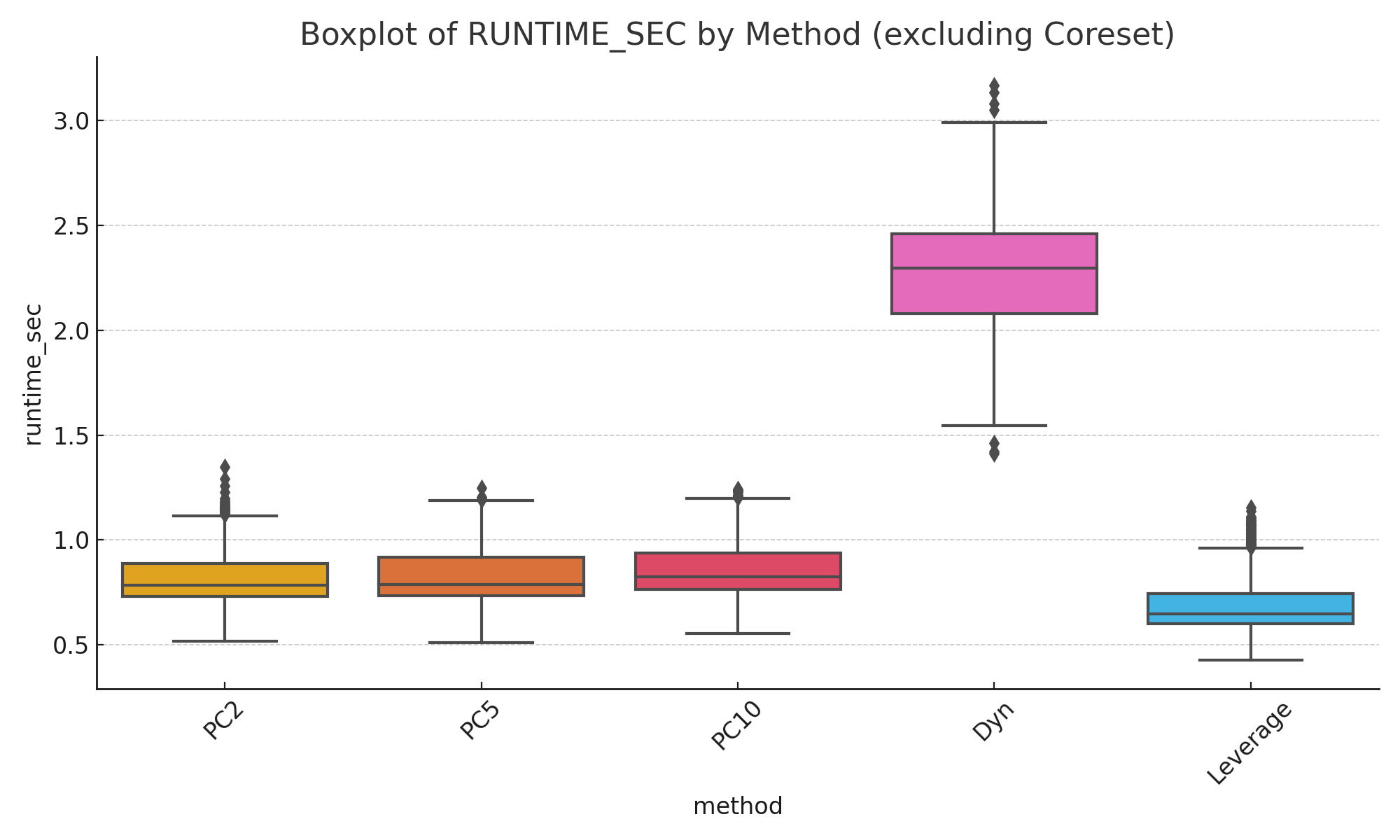}}
    \caption{Boxplots of model performance metrics across sampling methods. The Dynamic70 group is aggregated under the label \texttt{Dyn}, with the mean number of selected principal components = 266.25 (SD = 0.64). (Actual number of PCs used with dynamic pc option: mean: 266.25, standard deviation : 0.64, min: 264,'max: 268.)  Runtime is reported with and without inclusion of the Coreset method due to its distinct computational cost.}
    \label{fig:boxplots_all_metrics}
\end{figure}

Table~\ref{tab:performance_highlighted} and Figure~\ref{fig:boxplots_all_metrics} present a comprehensive performance comparison of linear regression models trained on PCA-QS retentive subsets, evaluated under varying PCA configurations, including fixed numbers of principal components (PC2, PC5, PC10), a dynamic selection strategy (Dyn), and two benchmark methods: Leverage Score and Coreset sampling. All simulations were conducted under a feature correlation setting of \( \rho = 0.2 \), with sample size \( n = 10000 \) and feature dimension \( p = 500 \).

\paragraph{Computational Efficiency}

The runtime results reveal that PCA-QS offers substantial computational savings relative to baseline methods. Specifically, the average CPU time for PCA-QS sampling ranges from 0.82 seconds (PC2) to 0.86 seconds (PC10), while the dynamic variant (Dyn) incurs slightly higher computation at 2.27 seconds due to adaptive PC determination. In contrast, Leverage Score sampling requires approximately 0.69 seconds, and Coreset sampling is significantly more costly, averaging 100.3 seconds per replication. Thus, PCA-QS—particularly in its fixed-PC configurations—offers a favorable trade-off between speed and representational quality, while avoiding the excessive runtime overhead associated with Coreset approaches.

\paragraph{Predictive Performance}

In terms of model accuracy, PCA-QS achieves low mean squared error (MSE) and high \( R^2 \) values across all configurations. Notably, the dynamic variant (Dyn) achieves the best performance among all methods, with an MSE of 0.238 and an \( R^2 \) of 0.953, closely approximating the full model performance while using only a subset of data. The Leverage and Coreset methods both exhibit higher MSEs (0.464 and 0.475, respectively) and lower \( R^2 \) values (0.908 and 0.905). Fixed-PC PCA-QS variants (PC2, PC5, PC10) show stable performance, with \( R^2 \) values ranging from 0.902 to 0.920. These results confirm that PCA-QS retains high predictive accuracy while substantially reducing computation time and data volume.

\begin{table}
\caption{%
Performance comparison of linear regression models trained on sampled subsets. 
Each cell shows the mean (standard deviation) over 1000 runs. 
\texttt{Dyn} (PCA-QS with dynamic PC selection) achieves the best accuracy. 
\texttt{Leverage} is fastest, while \texttt{Coreset} is the most computationally expensive.
}
\label{tab:performance_highlighted}
\begin{tabular}{lrrr}
\toprule
  Method &                     MSE &                       R2 &            Runtime (sec) \\
\midrule
     PC2 &         0.4674 (0.0219) &          0.9069 (0.0196) &           0.8192 (0.135) \\
     PC5 &         0.4912 (0.0252) &          0.9021 (0.0207) &          0.8267 (0.1383) \\
    PC10 &         0.4029 (0.0151) &          0.9197 (0.0167) &            0.86 (0.1415) \\
      Dyn & \textbf{0.238 (0.0034)} & \textbf{0.9526 (0.0097)} &           2.2668 (0.273) \\
Coreset &         0.4747 (0.0221) &          0.9054 (0.0198) &        100.3096 (5.3999) \\
Leverage &         0.4637 (0.0211) &          0.9076 (0.0193) & \textbf{0.6922 (0.1451)} \\
\bottomrule
\end{tabular}
In this table, Dyn denotes the case with PC numbers based on the PC variance loading 70\%.
Actual number of PCs used with dynamic pc option: mean: 266.248, standard deviation : 0.6395, min: 264,'max: 268.  
\end{table}

\paragraph{Computational Efficiency}
Relative CPU time remains lowest for PCA-QS with smaller numbers of components (PC2 and PC5) but increases as more components are included, reflecting additional computational cost. The Dynamic PCs approach requires more time due to the extra step of determining the optimal number of components, while the Coreset method incurs the highest runtime overall. Among all methods, Leverage sampling offers the fastest execution after PCA-QS with low PC counts.  A consolidated comparison of the theoretical properties and computational trade-offs of PCA-QS, Leverage, and Coreset sampling is provided in Appendix \ref{app:comparions}.

\paragraph{Optimal Number of Principal Components}
These findings suggest that using 2 to 5 principal components offers a strong balance between computational efficiency and predictive performance for PCA-QS. While adding more components (e.g., PC10 or Dynamic PCs) can yield marginal improvements in model accuracy, the computational cost rises noticeably. Overall, PCA-QS demonstrates significant gains in data reduction and runtime savings while maintaining high predictive accuracy, with configurations around 2–5 PCs appearing optimal for this simulation setting.

\begin{remark}
Selecting the number of principal components can be guided by the proportion of variance explained, commonly choosing a threshold such as 95\% based on conventional PCA analysis \citep{jolliffe2002principal, Jackson1991}. Alternatively, practical constraints such as computational cost or interpretability may dictate a smaller number of components. Extensive discussions on PCA variance loading and component selection can be found in the literature \citep{jolliffe2002principal, Abdi2010}. For clarity and simplification, this automatic selection step was not implemented in the present study.
\end{remark}

\paragraph{Large-Scale Linear Model Simulation.}
To examine the scalability of PCA-QS under larger data settings, we repeated the linear model simulation in Section~3.1.1 with an increased sample size of $n = 100{,}000$ while keeping all other parameters unchanged. The results confirm that PCA-QS maintains strong predictive performance and substantial reductions in data size and computational time even in this large-scale scenario.
\begin{table}[htbp]
\centering
\caption{Summary of performance metrics for different sampling methods under the large-scale linear model simulation ($n=100{,}000$).}
\label{tab:performance_rearranged_large}

\begin{subtable}[t]{0.9\textwidth}
\centering
\begin{tabular}{lcccc}
\toprule
Method & Energy & Mahalanobis & KL Divergence & MMD \\
\midrule
PCA-QS & 0.00005 ($\approx 0$) & 0.0154 (0.0001) & 0.00117 ($\approx 0$) & 0.0003 ($\approx 0$) \\
Leverage & 0.0789 (0.0084) & 0.5221 (0.0275) & 1.9505 (0.1267) & 0.4361 (0.0493) \\
Coreset & 0.2342 ($\approx 0$) & 0.8566 ($\approx 0$) & 3.587 ($\approx 0$) & 1.336 ($\approx 0$) \\
SRS & 0.0018 (0.0003) & 0.0911 (0.0100) & 0.3192 (0.1240) & 0.0084 (0.0022) \\
\bottomrule
\end{tabular}
\caption{Distance and divergence metrics}
\label{tab:subtable_distances}
\end{subtable}

\vspace{1em}

\begin{subtable}[t]{0.6\textwidth}
\centering
\begin{tabular}{lcc}
\toprule
Method & MSE & Runtime (sec) \\
\midrule
PCA-QS & 0.0720 ($\approx 0$) & 0.8615 (0.1757) \\
Leverage & 0.0730 (0.0003) & 0.1549 (0.2389) \\
Coreset & 0.0742 ($\approx 0$) & 27.4645 (8.9609) \\
SRS & 0.0729 (0.0002) & 0.0131 (0.0335) \\
\bottomrule
\end{tabular}
\caption{Performance metrics}
\label{tab:subtable_performance}
\end{subtable}

\end{table}

\begin{figure}[htbp]
    \centering
    % Row 1
    \begin{subfigure}[b]{0.45\textwidth}
        \includegraphics[width=\textwidth]{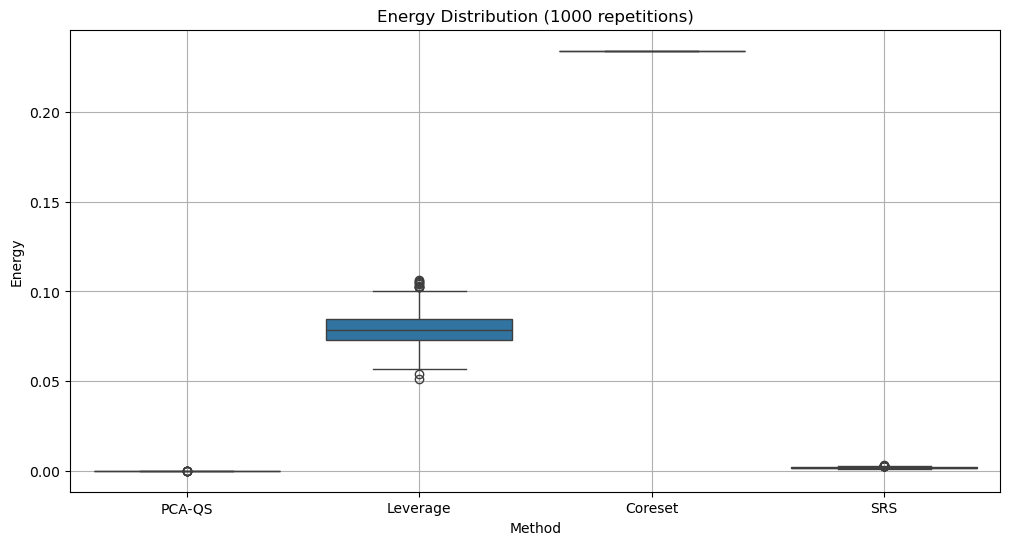}
        \caption{Energy Distribution}
    \end{subfigure}
    \hfill
    \begin{subfigure}[b]{0.45\textwidth}
        \includegraphics[width=\textwidth]{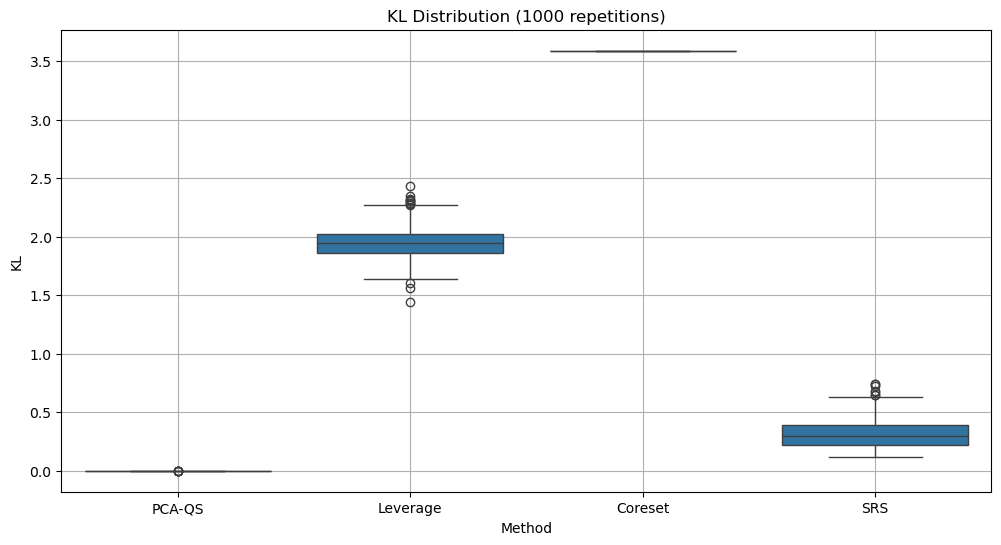}
        \caption{KL Divergence Distribution}
    \end{subfigure}
    
    % Row 2
    \vspace{0.5em}
    \begin{subfigure}[b]{0.45\textwidth}
        \includegraphics[width=\textwidth]{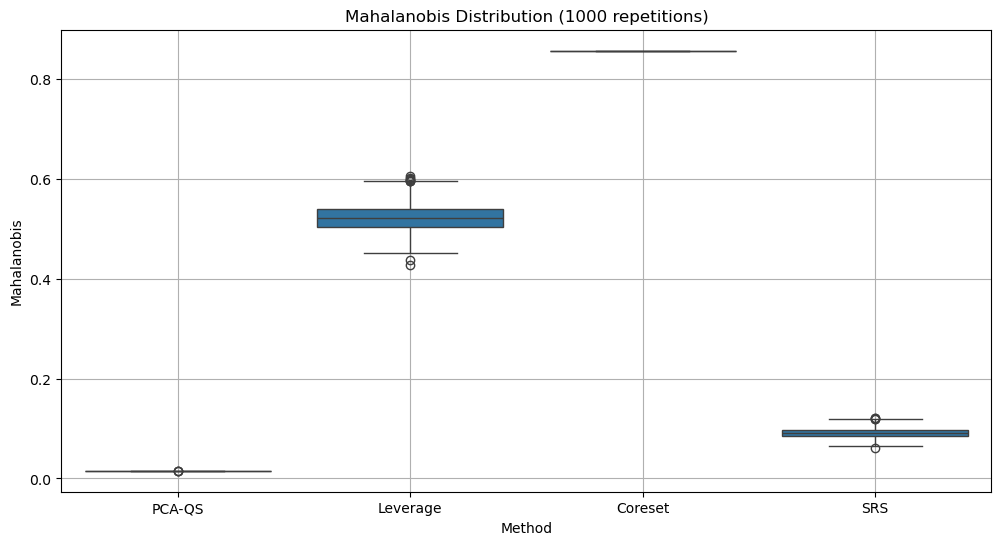}
        \caption{Mahalanobis Distance Distribution}
    \end{subfigure}
    \hfill
    \begin{subfigure}[b]{0.45\textwidth}
        \includegraphics[width=\textwidth]{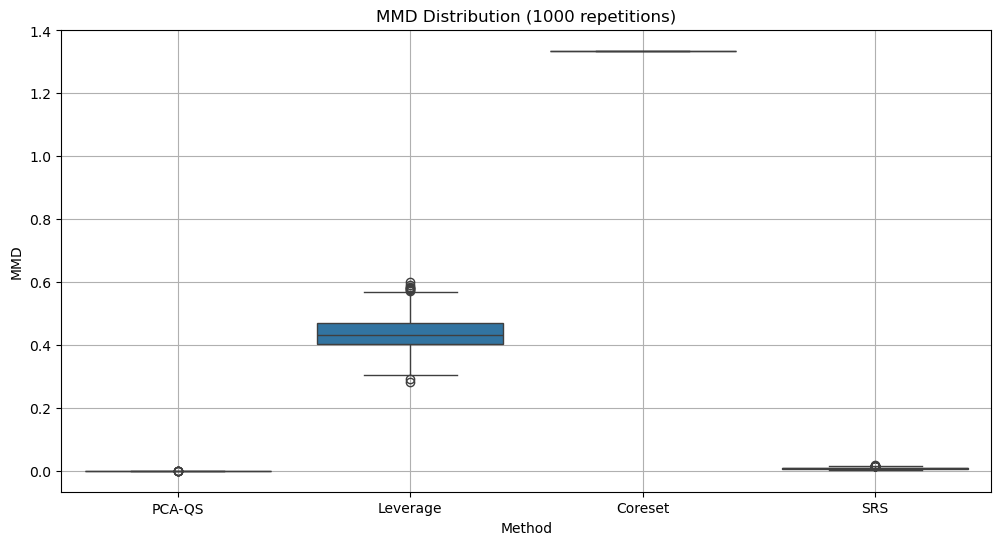}
        \caption{MMD Distribution}
    \end{subfigure}

    % Row 3
    \vspace{0.5em}
    \begin{subfigure}[b]{0.45\textwidth}
        \includegraphics[width=\textwidth]{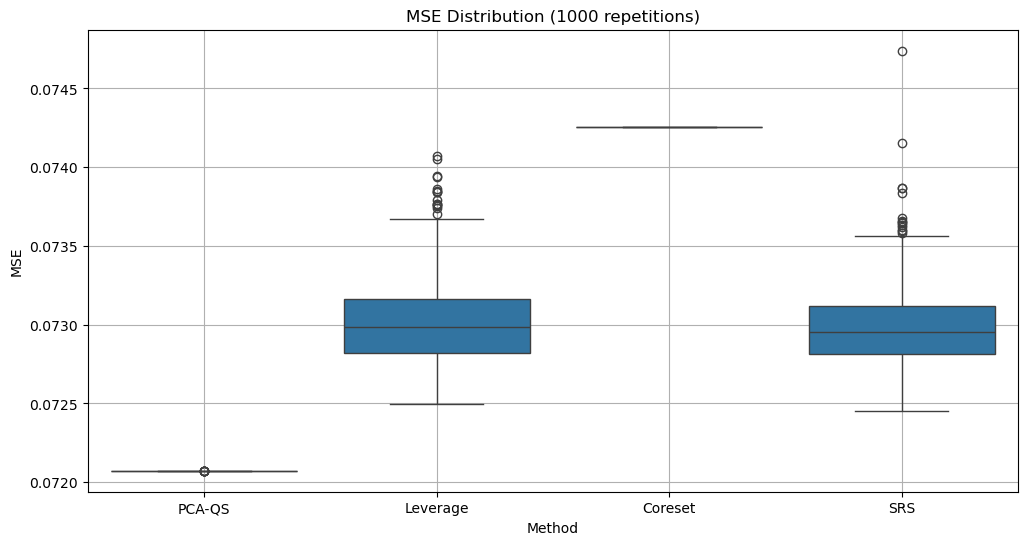}
        \caption{MSE Distribution}
    \end{subfigure}
    \hfill
    \begin{subfigure}[b]{0.45\textwidth}
        \includegraphics[width=\textwidth]{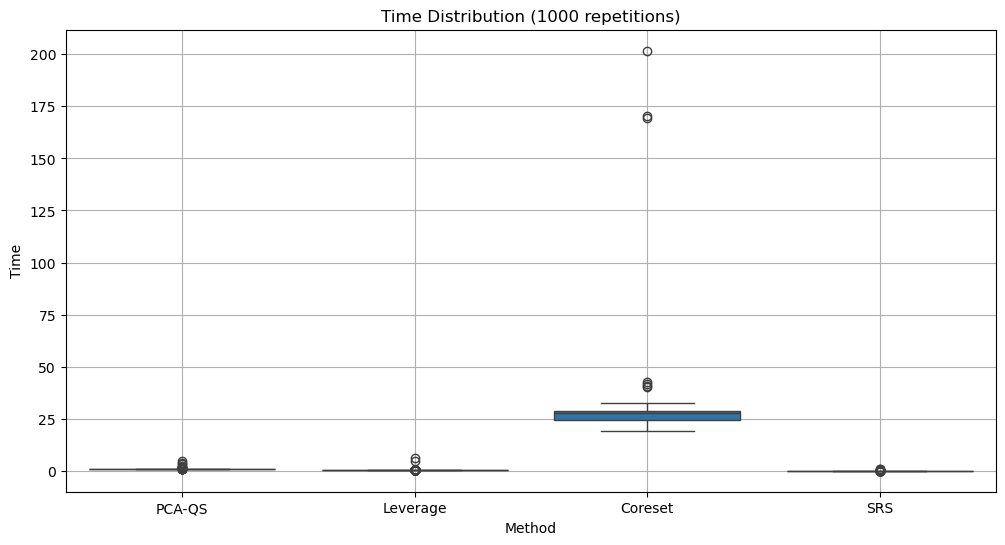}
        \caption{Time Distribution}
    \end{subfigure}
    
    \caption{Distributions of performance metrics across 1,000 repetitions for different sampling methods.}
    \label{fig:all_metrics_boxplots}
\end{figure}

\subsection{Application of PCA-QS to Clustering Problems}
In addition to the numerical studies presented in this paper, which focus on linear and logistic regression models as well as adaptive designs applied to various real-world datasets, this section explores the application of PCA-QS to clustering problems. While this is not an exhaustive investigation of clustering—given the inherent complexities of clustering methods—it offers valuable insights for future research. Specifically, PCA-QS is applied to the Cover Type dataset, a high-dimensional real-world dataset, to assess its utility as a dimensionality reduction technique for clustering tasks.

Dimensionality reduction techniques, particularly PCA, are widely studied for their ability to simplify data while preserving analytical accuracy \cite{jolliffe2002principal, abdi2010pca}. Clustering, however, remains inherently challenging due to the need for parameter selection, including determining the optimal number of clusters, choosing appropriate similarity measures, and managing sensitivity to noise and outliers \cite{hamerly2003learning, tibshirani2001gap}. Clustering performance is commonly evaluated using metrics such as the Silhouette Score, which measures cluster cohesion and separation \cite{rousseeuw1987silhouette}, and the Davies-Bouldin Index, which assesses compactness and separation \cite{davies1979cluster}. While robust and interpretable, these metrics often yield conflicting evaluations depending on the dataset and clustering algorithm used \cite{vendramin2010relative}.

By comparing clustering results between the retentive dataset (produced via PCA-QS) and the full dataset using k-means clustering, we observe highly similar outcomes, supporting the potential of PCA-QS to preserve essential data structure while improving computational efficiency and reducing noise. These findings align with existing research emphasizing the importance of maintaining structural integrity in reduced datasets for clustering tasks \cite{jain2010data, ding2004kmeans}. However, this study acknowledges the broader challenges of clustering analysis, particularly the difficulty in defining universally optimal parameters or performance criteria, which makes clustering highly context-dependent \cite{arbelaitz2013indices}. This addition broadens the scope of this paper by incorporating clustering applications and establishing a foundation for future research into the versatility of PCA-QS across diverse analytical paradigms.

\subsubsection{Clustering of Cover Type Data with the PCA-QS Method}

\subsubsection*{Cover Type Dataset Overview}
We also use the Cover Type dataset for illustration, despite its primary use in classification tasks. As mentioned earlier, it consists of cartographic attributes derived from remote sensing and soil characteristics to predict forest cover types across different land areas.
The target variable represents seven different forest cover types, such as Spruce/Fir, Lodgepole Pine, Ponderosa Pine, and Aspen.  Due to its high-dimensional nature and mix of categorical and continuous variables, the Cover Type dataset is frequently utilized in evaluating machine learning algorithms, clustering techniques, and feature selection methods.

\subsubsection*{Clustering Experiment Setup}
A clustering experiment was conducted on the Cover Type dataset using PCA-QS to extract a representative subset before clustering. To ensure robust evaluation, the experiment was repeated 1,000 times.

The clustering process followed three key steps:
\begin{itemize}
\item[]1. PCA-QS Sampling: A stratified sampling approach was applied, where 20\% of the dataset (deduction rate $\delta = 0.2$) was selected using PCA-QS. This ensured that the sampled data retained the distributional properties of the full dataset.
\item[]2. K-Means Clustering: The K-Means algorithm was applied to both the retentive (sampled) dataset and the full dataset, forming seven clusters corresponding to the seven forest cover types. The algorithm was initialized with 10 different random starts to enhance stability.
\item[]3. Performance Evaluation: Cluster assignments were predicted for the remaining data points using centroids from the retentive dataset and compared with those obtained from the full dataset. The clustering performance was evaluated using Silhouette Scores, computed separately for both sets using a parallelized silhouette computation method.
\end{itemize}
This approach ensured a scalable and efficient evaluation of clustering performance while assessing the effectiveness of PCA-QS in preserving the full dataset's structural integrity.

\subsubsection*{Clustering Performance and Results}
The absolute difference between the silhouette scores of the retentive and full datasets was less than 0.1 in 99.5\% of cases (995 out of 1,000). Under a stricter threshold of 0.05, 88.9\% of cases (889 out of 1,000) still met the criterion. This analysis demonstrates a high level of consistency between the two metrics, even under more stringent conditions.

\begin{figure}[h]
    \centering
    \includegraphics[width=\textwidth]{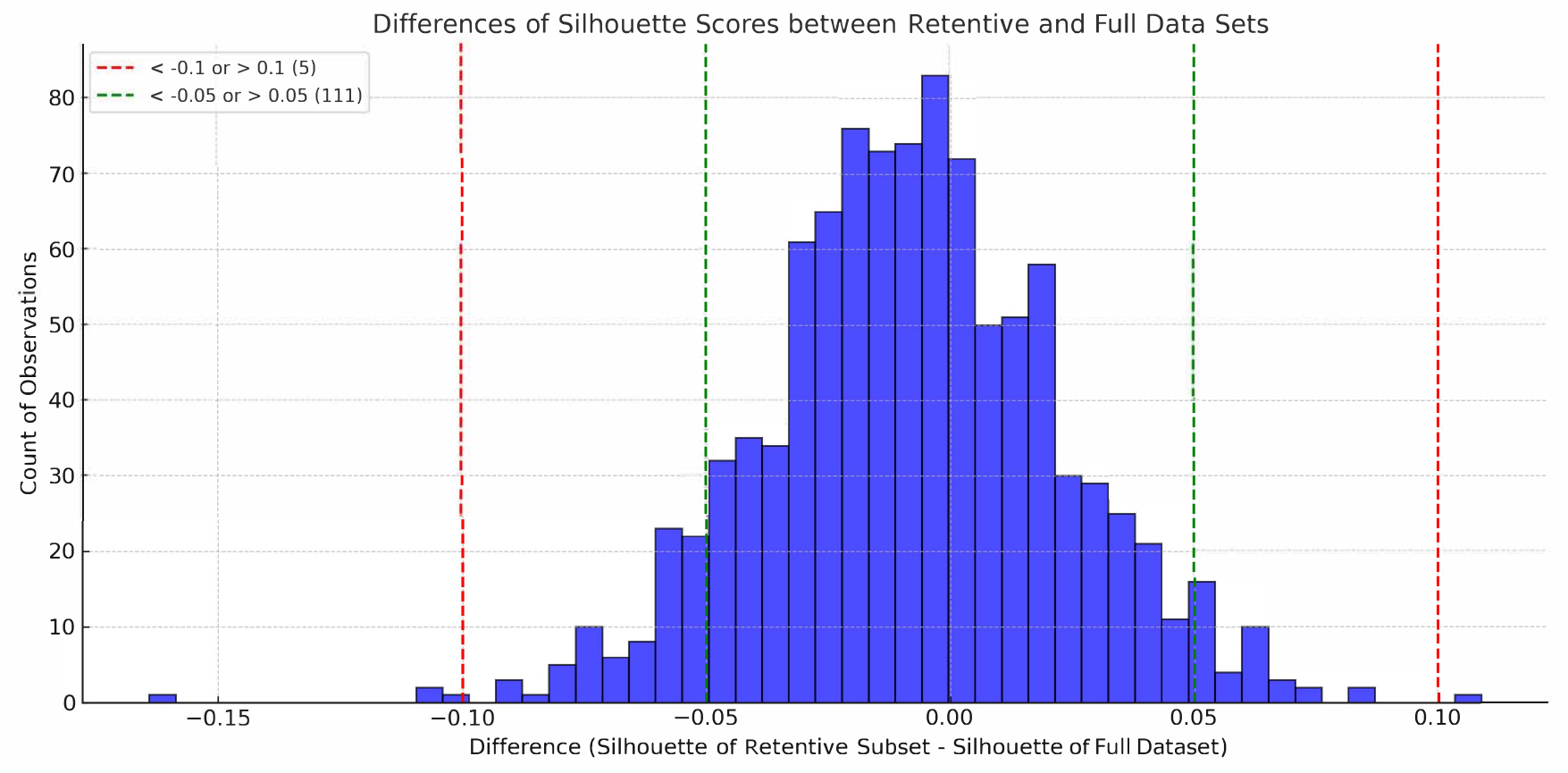}
   \caption{Distribution of Silhouette score of the differences between retentive and full datasets. 
The histogram illustrates the differences in silhouette scores between the retentive subset and the full dataset. 
Green dashed lines indicate the threshold at $\pm0.05$, covering 88.9\% of cases, while red dashed lines indicate the threshold at $\pm0.1$, covering 99.5\% of cases.}
    \label{fig:histogram_silhouette}
\end{figure}

Figure \ref{fig:histogram_silhouette} presents the distribution of differences between the Silhouette Score of the Retentive Dataset and the Silhouette Score of the Full Dataset across 1,000 observations. The x-axis represents the difference \((\text{Silhouette retentive dataset} - \text{Silhouette full dataset})\), while the y-axis denotes the count of observations.

The majority of differences are centered around zero, indicating strong agreement between the two metrics. Two threshold lines highlight key consistency ranges: 
\begin{itemize}
\item[] 1. Green dashed lines(\(\pm 0.05\)): 889 observations (\(88.9\%\)) fall within this range, while 111 exceed it.
\item[] 2. Red dashed lines (\(\pm 0.1\)): 995 observations (\(99.5\%\)) are within this threshold, with only 5 exceeding it.
\end{itemize}
The symmetrical distribution reinforces that deviations in either direction (positive or negative) are well-balanced. This visualization effectively illustrates the high level of similarity between the clustering results of the retentive and full datasets, with only a small fraction of observations displaying significant differences.

\paragraph{Impact of Data Deduction on Clustering Models}

The sensitivity of linear and clustering models to data deduction effects—such as missing values, outliers, and data reduction—varies significantly. Linear models, including logistic regression, are particularly affected by missing data, often requiring imputation to prevent biased parameter estimation \cite{little2002statistical}. Outliers can also distort predictions, necessitating robust regression techniques \cite{rousseeuw2005robust}. While PCA improves interpretability and mitigates multicollinearity, improper feature removal may reduce predictive power \cite{jolliffe2002principal}. 

Clustering models exhibit diverse sensitivities depending on the algorithm. Distance-based methods like K-Means struggle with missing values and high-dimensional spaces, often leading to inaccurate clustering \citep{hastie2009elements}. They are also susceptible to outliers, which distort cluster centroids, whereas density-based approaches like DBSCAN offer greater robustness \cite{ester1996density}. The curse of dimensionality further complicates clustering, often necessitating feature selection or structured sampling techniques to ensure representative instance selection.

Overall, clustering models tend to be more sensitive to data reduction effects than linear models, particularly in high-dimensional settings. This underscores the need for tailored preprocessing strategies when applying PCA-QS, reinforcing its utility in both regression and clustering contexts.

\subsection{PCA-QS With Other Experimental Design Based Adaptive Sampling Strategies}
\label{subsec:Adaptive sampling}
Adaptive sampling is widely used in modern big data analysis to efficiently locate informative data points within large datasets.
In previous numerical studies, the PCA-QS method was implemented with a random sampling scheme, where data points were randomly selected within quantile groups defined in PCA space. Since PCA-QS assigns PCA-quantile indexes to data points, the method can seamlessly incorporate other sampling strategies beyond randomness. In this section, we evaluate PCA-QS under adaptive sampling schemes.
 It follows an iterative process, where selected data points maximize information gain based on specific optimization criteria. We investigate whether PCA-QS can benefit from adaptive sampling techniques, particularly A-optimal, D-optimal, G-optimal design-based methods, and uncertainty sampling.

The data generation setup remains identical to the linear model case, where each sampling scheme is applied to datasets consisting of 100,000 samples and 500 variables. The results are based on 1,000 replications to ensure robustness. 
A brief introduction to each adaptive sampling method is provided in the following subsections, followed by numerical evaluations demonstrating how PCA-QS maintains computational efficiency and statistical integrity across different experimental design criteria.

{\color{black}
\paragraph{A-Optimal Design}
The A-optimal design minimizes the average variance of parameter estimates by reducing the trace of the inverse Fisher Information Matrix (FIM). This criterion enhances the precision of model parameter estimates on average, making it widely applicable in regression modeling and scenarios where accurate parameter estimation is a priority \citep{pukelsheim2006optimal, fedorov1972theory}.

Table \ref{tab:A-optimal} presents the computational and predictive performance of PCA-QS under the A-optimal scheme. PCA-QS achieves substantial computational savings, with CPU times for retentive datasets ranging from 0.0207 to 0.0195 seconds, compared to over 2.8 seconds for the full dataset. While the retentive datasets exhibit slightly higher MSE values (\(\approx 0.3126\)) relative to the full dataset (\(0.2498\)), the relative MSE remains stable at approximately 1.25, indicating that parameter estimation precision is well maintained. We use Ret and Full to denote the retentive subset and full dataset in tables below. 
Efficiency scores are consistently high, peaking at 2,423 for PCA = 2, suggesting that A-optimality within PCA-QS effectively balances computational efficiency and statistical accuracy.

\begin{table}[ht]
\caption{PCA-QS with A-optimal Design}
\label{tab:A-optimal}
\centering
\begin{tabular}{lrrrrrr}
  \hline
PCA & PCA  = 1 & PCA  = 2 & PCA  = 3 & PCA  = 4 & PCA  = 5 \\ 
  \hline
  CPU Ret & 0.0207 (0.0038) & 0.0196 (0.0026) & 0.0192 (0.0022) & 0.0195 (0.0021) & 0.0195 (0.0022) \\ 
  CPU Full & 2.8298 (0.4576) & 2.9187 (0.5514) & 2.7643 (0.4902) & 2.7573 (0.4783) & 2.8035 (0.5369) \\ \\ 
  MSE Ret  & 0.3126 (0.0087) & 0.3125 (0.0088) & 0.3125 (0.0088) & 0.3126 (0.0089) & 0.3126 (0.0089) \\ 
  MSE Full & 0.2498 (0.0011) & 0.2498 (0.0011) & 0.2498 (0.0011) & 0.2498 (0.0011) & 0.2498 (0.0011) \\ \\
  R Sq Ret  & 0.7408 (0.0266) & 0.7404 (0.0267) & 0.7403 (0.0266) & 0.7403 (0.0265) & 0.7403 (0.0265) \\ 
  R Sq  Full & 0.7928 (0.0217) & 0.7924 (0.0218) & 0.7923 (0.0217) & 0.7924 (0.0215) & 0.7924 (0.0215) \\ \\
  Rel MSE & 1.2516 (0.0329) & 1.2512 (0.0329) & 1.2511 (0.0332) & 1.2515 (0.0333) & 1.2515 (0.0334) \\ 
  Eff Score & 2229.26 (464.85) & 2423.26 (544.27) & 2332.71 (520.55) & 2292.76 (477.98) & 2330.95 (516.62) \\ 
   \hline
\end{tabular}
\end{table}

\paragraph{PCA-QS with D-optimal Design}
The D-optimal design maximizes the determinant of the Fisher Information Matrix (FIM), thereby maximizing the volume of the confidence ellipsoid for parameter estimates. This criterion is widely used in experimental design, particularly in fields such as pharmaceutical trials and engineering, where improving parameter estimation efficiency is critical \citep{fedorov1972theory, chaloner1995bayesian}.

Table \ref{tab:D-optimal} presents the performance of PCA-QS under the D-optimal design. PCA-QS maintains substantial computational efficiency, with CPU times for the retentive subset as low as 0.0201 seconds, compared to over 3 seconds for the full dataset. While MSE values for the retentive subset (~0.3125) remain slightly higher than those for the full dataset (0.2493), the relative MSE is stable at approximately 1.25, indicating consistent parameter estimation accuracy.
Efficiency scores peak at 2,581 for PCA = 3, demonstrating the scalability and robustness of PCA-QS under the D-optimal criterion.

\begin{table}[ht]
\caption{PCA-QS Performance with D-Optimal Scheme}
\label{tab:D-optimal}
%\fontsize{10}{12}\selectfont
\centering
\begin{tabular}{lrrrrr}
  \hline
PCA & PCA = 1 & PCA = 2 & PCA = 3 & PCA = 4 & PCA = 5 \\ 
  \hline
CPU Ret    & 0.0203 (0.0027) & 0.0217 (0.0056) & 0.0202 (0.0033) & 0.0201 (0.0029) & 0.0209 (0.0036) \\ 
CPU Full   & 2.7365 (0.4514) & 3.2923 (0.6328) & 3.1911 (0.6572) & 2.8885 (0.5314) & 3.1156 (0.5886) \\ \\
MSE Ret    & 0.3122 (0.0158) & 0.3126 (0.0165) & 0.3125 (0.0165) & 0.3125 (0.0164) & 0.3126 (0.0165) \\ 
MSE Full   & 0.2493 (0.0009) & 0.2493 (0.0008) & 0.2493 (0.0008) & 0.2493 (0.0008) & 0.2493 (0.0008) \\ \\
$R^2$ Ret  & 0.7508 (0.0205) & 0.7515 (0.0191) & 0.7516 (0.0191) & 0.7515 (0.0191) & 0.7515 (0.0191) \\ 
$R^2$ Full & 0.8007 (0.0164) & 0.8016 (0.0157) & 0.8016 (0.0157) & 0.8015 (0.0157) & 0.8016 (0.0157) \\ \\
Rel MSE    & 1.2521 (0.0629) & 1.2539 (0.0652) & 1.2539 (0.0653) & 1.2537 (0.0649) & 1.2540 (0.0654) \\ 
Eff Score  & 2183.67 (410.25) & 2503.78 (618.14) & 2581.74 (673.00) & 2336.44 (540.93) & 2446.10 (600.01) \\ 
  \hline
\end{tabular}
%Ret: Retentive subset; Full: Full dataset. 
\end{table}
}

\paragraph{PCA-QS with  G-Optimal Design}
G-optimal design minimizes the maximum variance of predicted values over the design space, ensuring uniformly distributed prediction errors. This approach is particularly useful in prediction-focused applications such as machine learning and response surface optimization \citep{pukelsheim2006optimal, cox1958planning}.

In PCA-QS, G-optimality reduces extreme variances while maintaining computational efficiency. Unlike other criteria that minimize average variance, G-optimal design ensures no single region dominates the prediction error, making it ideal for uniform performance across the design space.

Empirical results show that PCA-QS under G-optimality achieves substantial computational savings, with retentive subset CPU times between 0.0819 and 0.0203 seconds, compared to over 3 seconds for the full dataset (see Table \ref{tab:goptimal_pcaqs}). MSE values show slight variation across PCA configurations, with the highest values at PCA=3 and PCA=5 (approximately 0.3139). The relative MSE increase (around 1.26) aligns with the goal of reducing maximum variance rather than optimizing mean performance.
Efficiency scores peak at PCA=5 (approximately 2,521), indicating that higher PCA components offer better computational savings while maintaining variance control. These results suggest that PCA-QS effectively balances prediction uniformity and computational efficiency under G-optimal design.

\begin{table}[ht]
\centering
\caption{Performance of PCA-QS under G-optimal Design}
\label{tab:goptimal_pcaqs}
%\fontsize{10}{12}\selectfont
\begin{tabular}{lccccc}
  \hline
Metric & PCA 1 & PCA 2 & PCA 3 & PCA 4 & PCA 5 \\ 
  \hline
CPU Ret  & 0.0819 (0.0117) & 0.0203 (0.0032) & 0.0197 (0.0027) & 0.0212 (0.0042) & 0.0203 (0.0032) \\ 
CPU Full & 2.0987 (0.5749) & 2.9489 (0.4652) & 3.0283 (0.6183) & 3.0811 (0.5421) & 3.1424 (0.6203) \\ \\
MSE Ret  & 0.2539 (0.0018) & 0.3131 (0.0164) & 0.3139 (0.0173) & 0.3138 (0.0173) & 0.3139 (0.0174) \\ 
MSE Full & 0.2493 (0.0015) & 0.2491 (0.0007) & 0.2490 (0.0005) & 0.2490 (0.0005) & 0.2490 (0.0005) \\ \\
R Sq Ret  & 0.7926 (0.0241) & 0.7542 (0.0190) & 0.7547 (0.0177) & 0.7546 (0.0177) & 0.7546 (0.0176) \\ 
R Sq Full & 0.7964 (0.0236) & 0.8042 (0.0140) & 0.8051 (0.0129) & 0.8051 (0.0128) & 0.8052 (0.0128) \\ \\
Rel MSE & 1.0185 (0.0027) & 1.2567 (0.0644) & 1.2603 (0.0678) & 1.2600 (0.0677) & 1.2606 (0.0680) \\ 
Eff Score & 502.78 (110.61) & 2369.16 (511.72) & 2489.96 (632.38) & 2381.87 (596.49) & 2521.60 (638.61) \\ 
   \hline
\end{tabular}
%Ret: Retentive subset; Full: Full dataset. 
\end{table}

\paragraph{PCA-QS with Uncertainty-Based Design}
Uncertainty-based adaptive sampling dynamically updates the sampling strategy by incorporating model uncertainty and prior knowledge, reducing uncertainty iteratively. This approach is critical in fields such as environmental modeling, robotics, and Bayesian optimization \citep{chaloner1995bayesian, mackay1992information}.

Table \ref{tab:uncertainty} shows that under uncertainty-based sampling, PCA-QS maintains consistent computational efficiency, with retentive subset CPU times ranging from 0.0953 to 0.0967 seconds, compared to approximately 3 seconds for the full dataset. The retentive datasets exhibit minimal variation in MSE (~0.2541) across PCA configurations, closely aligning with the full dataset (~0.2493). Relative MSE remains stable (~1.02), and efficiency scores exceed 600 for higher PCA configurations, demonstrating the robustness of PCA-QS in uncertainty-aware settings.

\begin{table}[ht]
\caption{PCA-QS with Uncertainty-Based Sampling}
\label{tab:uncertainty}
%\fontsize{9}{11}\selectfont      
\centering
\begin{tabular}{lrrrrrr}
  \hline
PCA & PCA = 1 & PCA = 2 & PCA = 3 & PCA = 4 & PCA = 5 \\ 
  \hline
CPU Ret  & 0.0953 (0.0119) & 0.0966 (0.0174) & 0.0975 (0.0173) & 0.0966 (0.0186) & 0.0967 (0.0162) \\ 
CPU Full & 2.6915 (0.3769) & 2.9057 (0.5384) & 2.9637 (0.5448) & 3.0313 (0.5672) & 2.9112 (0.4859) \\ \\
MSE Ret  & 0.2541 (0.0015) & 0.2541 (0.0015) & 0.2541 (0.0015) & 0.2541 (0.0015) & 0.2541 (0.0015) \\ 
MSE Full & 0.2493 (0.0011) & 0.2493 (0.0011) & 0.2493 (0.0011) & 0.2493 (0.0011) & 0.2493 (0.0011) \\ \\
R Sq Ret  & 0.7833 (0.0188) & 0.7829 (0.0187) & 0.7829 (0.0187) & 0.7829 (0.0187) & 0.7828 (0.0187) \\ 
R Sq Full & 0.7874 (0.0184) & 0.7870 (0.0184) & 0.7870 (0.0183) & 0.7870 (0.0183) & 0.7869 (0.0183) \\ \\
Rel MSE & 1.0194 (0.0026) & 1.0194 (0.0026) & 1.0194 (0.0026) & 1.0194 (0.0026) & 1.0194 (0.0026) \\ 
Eff Score & 560.58 (95.88) & 604.98 (142.98) & 611.33 (146.19) & 634.99 (159.49) & 604.45 (135.37) \\ 
   \hline
\end{tabular} 
%Ret: Retentive subset; Full: Full dataset. 
\end{table}

%\subsection*{Advantages of Quantile Grouping in PCA-QS for Adaptive Designs}

It follows from above experiments,  we can see that across all adaptive sampling schemes, PCA-QS consistently reduces computational costs while preserving statistical integrity. The method adapts effectively to various design criteria, including variance minimization, determinant maximization, and uncertainty-based strategies. Efficiency scores and relative MSE values demonstrate its scalability and versatility, making PCA-QS a practical tool for high-dimensional data reduction across diverse modeling scenarios.

Quantile grouping in PCA-QS provides significant computational and statistical advantages, particularly in adaptive sampling designs based on statistical experimental design criteria. Since these methods evaluate each sample within predefined groups, quantile grouping ensures representativity, preserves data heterogeneity, and enhances model robustness. By stratifying data into quantile-based groups, computational workload is efficiently distributed, reducing redundant calculations and improving scalability. This structured approach facilitates systematic exploration of the feature space, leading to greater estimation accuracy and stability in iterative sampling and decision-making processes.

Balanced sample allocation within each quantile group allows adaptive algorithms to refine experimental design criteria effectively, reducing bias and improving parameter estimation and predictive performance. PCA-QS validation on high-dimensional linear regression models (500 features, 100,000 samples) confirms no significant differences in covariance and response vectors between retentive and full datasets, as verified by Kolmogorov-Smirnov and T-tests. The method is well-suited for broader applications.

Furthermore, PCA-QS is highly compatible with parallel and distributed computing. By grouping data in the PCA-transformed space, it enables independent sampling within each group, significantly reducing computational complexity in adaptive sampling schemes such as A- and D-optimality.

\subsection{Similarity Analysis on Real-World Datasets}
The primary goal of the following analysis is to examine how well each sampling method preserves the statistical distribution of the original data after reduction. Thus, in addition to synthetic data experiments, we evaluate the performance of PCA-QS, Leverage Score Sampling, Coreset Sampling, and SRS on several real-world datasets from the UCI Machine Learning Repository.  We assess distributional similarity using multiple distance metrics, including the Energy distance, Mahalanobis distance, Kullback--Leibler (KL) divergence, and Maximum Mean Discrepancy (MMD. These metrics quantify various aspects of how closely the retentive datasets approximate the original data distribution in both global structure and local variability.
The datasets selected for this study span a range of domains and dimensionalities:
\begin{itemize}
    \item[] \textbf{CAPS} Dataset: A collection of physiological and behavioral measurements used in mental health research.
    \item[] \textbf{Online News Popularity}: Features related to online articles and their social media popularity.
    \item[] \textbf{Bike Sharing Dataset (hours)}: Hourly records of bike rentals, including temporal, weather, and holiday features.
    \item[] \textbf{Application Prediction Dataset}: Demographic and behavioral variables predicting app installation or usage.
\end{itemize}
For each dataset, we apply a uniform retention rate of $10\%$, repeating all experiments 1,000 times and reporting the mean and standard deviation for each metric. Table~\ref{tab:real_data_similarity_time} presents the results for CAPS, Online News Popularity, Bike Sharing, and Appliance Energy. PCA-QS and SRS generally achieve the lowest values across most metrics, with SRS performing best on Bike Sharing and Appliance Energy, and PCA-QS excelling on CAPS and Online News Popularity. The smallest (best) values for each distance metric are highlighted in bold. Results for leverage score and coreset sampling methods vary by dataset. Although execution time is not reported, SRS is the fastest among all four methods, with PCA-QS also typically more efficient than the remaining approaches. Note that for the results shown in this table, PCA-QS uses a dynamic number of principal components, retaining no less than $70\%$ of the total variance for each dataset.
\begin{table}[ht]
\centering
%\small
\caption{Mean and standard deviation of the distributional similarity metrics and runtime between the retentive subsets and the original data across real-world datasets based on 1000 repeated runs.}
\label{tab:real_data_similarity_time}
\begin{tabular}{llllll}
\hline
\textbf{Dataset} & \textbf{Method} & \textbf{Energy} & \textbf{Mahalanobis} & \textbf{KL} & \textbf{MMD}\\ % & \textbf{Time (s)} \\
\hline
\multirow{4}{*}{CAPS}
& PCA-QS   & {\bf 0.013} (0.007) & {\bf 0.032} (0.0087) & 0.157 (0.258) & {\bf 0.0004} (0.0003) \\ %& 0.539 (13.516) \\
& Leverage & 0.485 (0.102) & 0.773 (0.0161) & 1.666 (0.022) & 2.437 (0.120) \\ %& 0.055 (0.034) \\
& Coreset  & 0.367 (0.083) & 0.766 ($\approx$ 0) & {\bf 1.402} ($\approx$ 0) & 1.897 ($\approx$ 0) \\ %& 81.716 (220.277) \\
& SRS      & 0.014 (0.008) & 0.041 (0.0099) & 0.164 (0.248) & 0.0016 (0.0016) \\ %& {\bf 0.004} (0.004) \\
\hline
\multirow{4}{*}{Online}
& PCA-QS   & 1.428 (3.604) & {\bf 0.040} (0.006) &  {\bf 0.001} ($\approx$ 0) & 8 $\times 10^{-6}$ ($\approx$ 0) \\ %& 2$\times 10^{-6}$ ($\approx$ 0) \\
& Leverage & 1.233 (30.679) & 0.203 (0.0366) & 0.756 (0.016) & 5.131 (0.101)\\ % & 1.329 (0.065) \\
& Coreset  & 240.79 (350.92) & 0.153 (0.029) & 0.655 ($\approx$ 0) & 5.403 ( $\approx$ 0)\\ % & 1.174 ($\approx$ 0) \\
& SRS      & {\bf 0.016} (0.023) & {\bf 0.040} (0.006) & 0.112 (0.011) & 3$\times 10^{5}$ (2.276$\times 10^{5}$) \\ %& 0.013 (0.004) \\
\hline
\multirow{4}{*}{Bike Sharing}
& PCA-QS   & {0.023} (0.007)        & 0.096 ($4.9 \times 10^{-4}$) & {\bf 0.015} ($2.8 \times 10^{-4}$) & 0.015 ($1.4 \times 10^{-4}$) \\
& Leverage & 0.087 (0.020)        		& 0.513 (0.031)                   & 0.441 (0.036)                                   & 0.329 (0.040) \\
& Coreset  & 0.036 (0.009)        		& 0.274 ($\approx$ 0)      & 0.213 ($\approx$ 0).                       & 0.087 ($\approx$ 0) \\
& SRS      & {\bf 0.020} (0.005)  		& {\bf 0.084} (0.016)         & {0.055} (0.025)                              & {\bf 0.007} (0.003) \\
\hline
\multirow{4}{*}{Appliance}
& PCA-QS   & 0.0029 (0.0009) & 0.106 (0.0155) & 0.294 (0.0143) & $\approx$ 0 ($\approx$ 0) \\
& Leverage         & 0.0240 (0.0043) & 0.276 (0.0189) & 0.483 (0.0335) &$\approx$ 0 ($\approx$ 0) \\
& Coreset          & 0.0085 (0.0007) & 0.208 (0.0077) & 0.382 (0.0146) & $\approx$ 0 ($\approx$ 0) \\
& SRS              & 0.0032 (0.0011) & 0.108 (0.0152) & 0.299 (0.0151) & $\approx$ 0 ($\approx$ 0) \\
\hline
\end{tabular}
``Online'' and  ``Appliance'' denote the Online News Popularity and Appliance Energy datasets, respectively. There are two subsets of Bike Sharing -- hour and day, here are results for hours. Execution time is not included in this table. However, as expected, SRS requires less time than all other methods, and PCA-QS generally takes less time than Coreset and leverage score sampling methods. Note that PCA-QS here is based on dynamic principal components (PCs), so the number of PCs used varies among datasets.
\end{table}

Across all datasets, PCA-QS demonstrates consistently lower distributional distances compared to other methods, confirming its capability to preserve the statistical characteristics of the original data while achieving substantial data reduction. While Leverage Score Sampling achieves strong performance in certain modeling tasks, it often exhibits larger distributional distortions, reflected in higher distances. Coreset Sampling achieves low mean squared error but shows limitations in preserving local variation, as indicated by elevated Energy and MMD scores. SRS provides a fast and unbiased baseline but does not exploit structural information, leading to moderate performance across all metrics. 

These results reinforce PCA-QS as an effective tool for real-world data reduction tasks where maintaining distributional similarity and computational efficiency are both critical for downstream analyses.

\section{Conclusion}
\label{sec:conc}

This study presents the PCA-QS framework, a hybrid method designed to achieve efficient data reduction while preserving the essential distributional characteristics of the original dataset. Unlike conventional dimensionality reduction or random sampling methods, PCA-QS uses the structure revealed by PCA to guide stratified quantile-based selection, ensuring that retained subsets maintain global data representativity.

Extensive numerical experiments on synthetic mixture Gaussian models and various benchmark datasets demonstrate that PCA-QS consistently outperforms standard sampling techniques such as SRS, Leverage Score Sampling, and Coreset Sampling in preserving distributional similarity. Metrics including Energy distance, Mahalanobis distance, Kullback–Leibler divergence, and Maximum Mean Discrepancy confirm that PCA-QS achieves closer alignment between the reduced and original datasets compared to competing methods, while maintaining reasonable computational cost.  PCA-QS achieves low distortion between the retentive and original datasets, as measured by multiple distributional similarity metrics.

Furthermore, the integration of adaptive sampling strategies—such as A-optimal, D-optimal, G-optimal, and uncertainty-based selection—demonstrates that PCA-QS remains robust and versatile under various experimental design objectives. These extensions further enhance its ability to produce informative, representative subsets tailored for diverse analytical tasks beyond regression, including clustering and exploratory analysis.

While PCA-QS strikes a practical balance between data reduction and structural fidelity, some limitations persist. Its reliance on linear PCA may limit its effectiveness for highly non-linear manifolds or data with strong local dependencies. Additionally, quantile stratification introduces an exponential increase in group combinations as more principal components are retained, which may raise computational overhead for very high-dimensional datasets.
Future research could address these challenges by investigating non-linear extensions of the PCA step, hybrid sampling techniques that combine multiple stratification bases, or more scalable quantile assignment methods for massive datasets. 

PCA-QS offers a robust and interpretable framework for producing retentive subsets that faithfully mimic the original data distribution. This makes it a valuable tool for large-scale data reduction where maintaining distributional similarity is crucial for subsequent statistical modeling, visualization, or machine learning workflows. Its adaptive extensions further broaden its applicability, providing a flexible basis for future developments in efficient and representative sampling.

\begin{appendices}

\section{Further details regarding usages of PCA-QS, Leverage Score Sampling, Coreset Sampling and SRS}\label{secA1}
We  summarize the features of PCA-QS, SRS, Leverage Score Sampling and Coreset sampling in table below, with some practical recommendation about their usages.

\subsection{PCA-QS}

PCA-QS shows strong all-round performance, striking a balance between preserving the overall distribution and ensuring low model error. It achieves one of the lowest energy distances, Mahalanobis distances, KL divergence, and MMD, indicating that its selected subset closely resembles the original data distribution in both mean and variance structure. Its MSE is marginally higher than Leverage and Coreset, but this trade-off yields much better distributional fidelity. The runtime is moderate, as it includes PCA decomposition and stratified sampling.

\textit{When to use:} Recommended when balanced preservation of global data structure and interpretability are critical, e.g., for exploratory analysis, visualization, or general-purpose data reduction.

\subsection{Leverage Score Sampling}

Leverage score sampling attains the lowest MSE after Coreset and retains low computational cost due to efficient randomized SVD. However, it performs poorly on distributional similarity metrics: it drastically distorts the data distribution (high Energy, Mahalanobis, KL, and MMD). This happens because it preferentially samples high-influence rows, which skews the sample toward extreme points.

\textit{When to use:} Best suited when the goal is to maintain predictive accuracy in linear models, not to replicate the entire data distribution. A pragmatic choice for fast regression tasks with large datasets.

\subsection{Coreset Sampling}

Coreset sampling, approximated via K-Means cluster centers, achieves the lowest MSE with zero variance across replicates, highlighting its stability in summarizing cluster means. However, it loses local variation within clusters, which is reflected in higher Energy distance, Mahalanobis distance, KL divergence, and particularly elevated MMD. Its computational cost is the highest due to the iterative clustering procedure.

\textit{When to use:} Ideal when the primary goal is to approximate the data for clustering or centroid-based downstream tasks where preserving the fine-grained distribution is less critical. Useful when high accuracy is more important than fine data structure and when runtime is acceptable.

\subsection{Simple Random Sampling (SRS)}
SRS offers a fast, unbiased baseline that yields reasonable performance across all metrics. Its MSE is slightly higher than Leverage and Coreset, but it achieves respectable Energy, Mahalanobis, and KL scores, with moderate MMD. Its computational time is the lowest since it only performs random index selection.
\textit{When to use:} SRS is a practical baseline for small to moderate datasets or when no model structure information is available. It is a safe choice when computational simplicity is paramount
.
\subsection{Comparison of Sampling Methods: Coreset, Leverage Scores, and PCA-QS}
\label{app:comparions}

\paragraph{Coreset Sampling.}
Coreset sampling aims to construct a small, weighted subset of the data that preserves geometric and statistical properties of the original dataset \cite{feldman2011, bachem2018scalable}. A common strategy involves clustering the data into $k$ clusters and selecting representative samples from each. This approach is particularly effective for large sample sizes ($n$) because its computational complexity scales linearly with $n$ and the feature dimension $d$, with a runtime of approximately $\mathcal{O}(n \cdot d \cdot k)$. However, Coresets can struggle in very high-dimensional spaces where the notion of distance becomes less meaningful due to the curse of dimensionality \cite{aggarwal2001surprising}. Furthermore, when the sample size $n$ is small, forming meaningful clusters can be problematic as clusters may consist of only one or very few data points. Coreset sampling tends to be robust against outliers since clustering suppresses the impact of isolated points.

\paragraph{Leverage Score Sampling.}
Leverage score sampling is a probabilistic technique that samples rows from the data matrix based on their influence in linear models \cite{drineas2012fast, mahoney2011randomized}. The leverage score of a data point measures its contribution to the least-squares fit and is computed from the squared row norms of the top left singular vectors of the data matrix. Formally, for a data matrix $X \in \mathbb{R}^{n \times d}$ with singular value decomposition $X = U \Sigma V^\top$, the leverage score for the $i$-th row is $\|\mathbf{u}_i\|^2$, where $\mathbf{u}_i$ denotes the $i$-th row of $U$. Leverage sampling is theoretically optimal for linear regression and low-rank approximation \cite{drineas2006sampling}. It performs well even in high dimensions because it captures dominant variance directions rather than relying purely on distances. However, computing leverage scores can be computationally expensive, requiring $\mathcal{O}(n \cdot d^2)$ time for exact SVD or $\mathcal{O}(n \cdot d \cdot k)$ for randomized methods \cite{halko2011finding}. Moreover, leverage sampling can be sensitive to outliers, as outlying points often have disproportionately high leverage scores.

\paragraph{PCA-QS Sampling.}
PCA-QS is a hybrid approach that combines dimensionality reduction and stratified sampling along principal components. First, PCA reduces the data from $d$ dimensions to $k$ principal components, retaining most of the variance \cite{jolliffe2002principal}. Subsequently, quantiles along each selected principal component are computed, partitioning the low-dimensional space into regions. Sampling then proceeds by selecting points from each quantile region, ensuring broad coverage of data variability. PCA-QS excels in maintaining diversity across the dominant variance directions, offering a balance between purely geometric and purely probabilistic sampling methods. Its runtime is dominated by the PCA computation, costing approximately $\mathcal{O}(n \cdot d \cdot k)$ for randomized algorithms \cite{halko2011finding}. However, PCA-QS may be sensitive to the chosen number of principal components and quantile thresholds. Like leverage sampling, it can suffer when sample sizes are very small, as PCA may become unstable due to insufficient data.

\paragraph{Effect of Sample Size and Dimensionality.}
The performance of these methods varies substantially with the sample size $n$ and dimensionality $d$. Coreset sampling scales well with large $n$ but struggles with high $d$. Leverage sampling remains effective in high dimensions but incurs significant computational cost. PCA-QS offers a trade-off, providing effective coverage of the data distribution while mitigating some of the computational burdens of full leverage-score computation. For small $n$, both leverage sampling and PCA-QS can remain useful, while coresets may fail to form meaningful partitions.

In practice, Coresets are preferable for purely geometric or clustering applications, leverage sampling is optimal for regression tasks, and PCA-QS is an attractive compromise for maintaining diversity in the data representation while controlling computational costs. Table~\ref{tab:sampling_comparison} summarizes their comparative strengths and limitations.

\begin{table}[h]
\centering
\caption{Comparison of Coreset Sampling, Leverage Score Sampling, and PCA-QS}
\label{tab:sampling_comparison}
\begin{tabular}{|p{1.5cm}|p{3cm}|p{3cm}|p{3cm}|}
\hline
 & \textbf{Coreset} & \textbf{Leverage Score} & \textbf{PCA-QS} \\ \hline
\textbf{Large $n$} & Scales efficiently, good clustering & Expensive but effective &  Scales well with $n$ \\ \hline
\textbf{Small $n$} & Clusters become unreliable & Still works well for regression & Unstable if too few samples \\ \hline
\textbf{High $d$} & Clustering unreliable due to curse of dimensionality & Remains effective but costly & Effectively reduces dimension \\ \hline
\textbf{Low $d$} & Performs well & Fast computation & Fast and stable \\ \hline
%\textbf{Computational Cost} & $\mathcal{O}(n \cdot d \cdot k)$ & $\mathcal{O}(n \cdot d \cdot k)$ randomized SVD & $\mathcal{O}(n \cdot d \cdot k)$ randomized PCA \\ \hline
\textbf{Outlier Sensitivity} & Generally robust & Sensitive to high-leverage outliers & Sensitive if outliers dominate principal components \\ \hline
\end{tabular}
\end{table}




\end{appendices}


\bibliography{PCAQS_references}% common bib file

\end{document}